\documentclass[aps,prd,superscriptaddress,nofootinbib,tighten,preprint]{revtex4}
\pdfoutput=1
\usepackage[utf8]{inputenc}
\usepackage[english]{babel}
\usepackage{amsmath}
\usepackage{subfigure}
\usepackage{amsfonts}
\usepackage{amssymb}
\usepackage{graphicx}
\newcommand{\be}{\begin{equation}}
\newcommand{\ee}{\end{equation}}
\newcommand{\bea}{\begin{eqnarray}}
\newcommand{\eea}{\end{eqnarray}}

\newcommand{\bl}{U(1)_{\rm B-L}}
\newcommand{\vbl}{v_{BL}}
\newcommand{\nn}{\nonumber}

\catcode`@=12
\def\la{\mathrel{\mathchoice {\vcenter{\offinterlineskip\halign{\hfil
$\displaystyle##$\hfil\cr<\cr\sim\cr}}}
{\vcenter{\offinterlineskip\halign{\hfil$\textstyle##$\hfil\cr<\cr\sim\cr}}}
{\vcenter{\offinterlineskip\halign{\hfil$\scriptstyle##$\hfil\cr<\cr\sim\cr}}}
{\vcenter{\offinterlineskip\halign{\hfil$\scriptscriptstyle##$\hfil\cr<\cr\sim
\cr}}}}}
\def\ga{\mathrel{\mathchoice {\vcenter{\offinterlineskip\halign{\hfil
$\displaystyle##$\hfil\cr>\cr\sim\cr}}}
{\vcenter{\offinterlineskip\halign{\hfil$\textstyle##$\hfil\cr>\cr\sim\cr}}}
{\vcenter{\offinterlineskip\halign{\hfil$\scriptstyle##$\hfil\cr>\cr\sim\cr}}}
{\vcenter{\offinterlineskip\halign{\hfil$\scriptscriptstyle##$\hfil\cr>\cr\sim
\cr}}}}}
\begin{document}

\title{Galactic Gamma Ray Excess and Dark Matter Phenomenology
in a $U(1)_{B-L}$ Model}

\author{Anirban Biswas}
\email{anirbanbiswas@hri.res.in}
\affiliation{Harish-Chandra Research Institute, Chhatnag Road,
Jhunsi, Allahabad 211 019, India}
\author{Sandhya Choubey}
\email{sandhya@hri.res.in}
\affiliation{Harish-Chandra Research Institute, Chhatnag Road,
Jhunsi, Allahabad 211 019, India}
\affiliation{Department of Theoretical Physics, School of
Engineering Sciences, KTH Royal Institute of Technology, AlbaNova
University Center, 106 91 Stockholm, Sweden}
\author{Sarif Khan}
\email{sarifkhan@hri.res.in}
\affiliation{Harish-Chandra Research Institute, Chhatnag Road,
Jhunsi, Allahabad 211 019, India}

\begin{abstract}
In this work, we have considered a gauged $U(1)_{\rm B-L}$ extension
of the Standard Model (SM) with three right handed neutrinos for anomaly
cancellation and two additional SM singlet complex scalars with
nontrivial B-L charges. One of these is used to spontaneously
break the $U(1)_{\rm B-L}$ gauge symmetry, leading to Majorana masses
for the neutrinos through the standard Type I seesaw mechanism,
while the other becomes the dark matter (DM) candidate in the model.
We test the viability of the model to simultaneously explain the DM relic
density observed in the CMB data as well as the Galactic Centre
(GC) $\gamma$-ray excess seen by Fermi-LAT. We show that for DM masses
in the range 40-55 GeV and for a wide range of  $U(1)_{\rm B-L}$
gauge boson masses, one can satisfy both these constraints if the
additional neutral Higgs scalar has a mass around the resonance
region. In studying the dark matter phenomenology and GC excess,
we have taken into account theoretical as well as experimental
constraints coming from vacuum stability condition, Planck bound
on DM relic density, LHC and LUX and present allowed areas in
the model parameter space consistent with all relevant data,
calculate the predicted gamma ray flux from the GC and discuss
the related phenomenology. 

\end{abstract}
\maketitle

\section{Introduction}
\label{Intro}
The presence of dark matter in the Universe is now well
established. Its presence has been probed
by its gravitational interaction with the visible world such as
the rotation curve of spiral galaxies \cite{Sofue:2000jx},
gravitational lensing \cite{Bartelmann:1999yn}, and the 
phenomena of the Bullet cluster \cite{Clowe:2003tk}.
Meanwhile, the amount of dark matter present in the Universe
has already been measured with an unprecedented
accuracy by various satellite borne experiments
like WMAP \cite{Hinshaw:2012aka} and
Planck \cite{Ade:2015xua}. The results of these experiments reveal
that our Universe has more than 80\% of its matter content in the form of
dark matter while the remaining part is composed of known baryonic matter. 
However, the possible nature of the constituents
of dark matter and their interactions with the Standard Model (SM) particles
as well as with themselves still remains an enigma. As the Standard Model
of particle physics does not have any fundamental particle which can
play the role of a dark matter candidate, there exist many extensions of SM
which can accommodate a single \cite{Jungman:1995df, Bertone:2004pz,
McDonald:1993ex, Burgess:2000yq, Biswas:2011td, Barbieri:2006dq,
LopezHonorez:2006gr, Kim:2008pp, Kim:2016csm, Hambye:2008bq, Lebedev:2011iq}
or multicomponent dark matter scenarios
\cite{Cao:2007fy, Zurek:2008qg, Belanger:2011ww, Biswas:2013nn, Bian:2013wna,
Bhattacharya:2013hva}. Out
of these different types of dark matter scenarios the most favourable one
is the class of particles known as weakly interacting massive particle or WIMP
\cite{Gondolo:1990dk, Srednicki:1988ce}.
This class of dark matter particles were produced thermally at early stage of
the Universe and maintained their thermal as well as chemical equilibrium
through the interactions with other particles within the thermal plasma.
As the Universe expanded and cooled down their rates began to decrease and
eventually, when the annihilation rate of WIMP became less
than the expansion rate of the Universe, WIMP decoupled from
the thermal plasma and froze to a particular relic density.   

Particle nature of WIMP can be explored mainly in two possible ways.
First is the method of direct detection, where the information about
WIMP mass and the nature of its interaction with SM particles can be obtained 
by measuring the recoil energy of the detector nuclei scattered by the
WIMP. There are many ongoing dark matter direct detection experiments 
such as LUX \cite{Akerib:2013tjd, Akerib:2015rjg}, XENON1T
\cite{Aprile:2015uzo} and SuperCDMS \cite{Agnese:2014aze}. However,
none of them have observed yet any ``real event'' which is produced by
the scattering of dark matter particles with the detector nuclei and have thus
placed an upper bound on both the spin independent and spin dependent
scattering cross sections of WIMP as a function of its mass.  
Another promising method is the indirect detection of dark matter,
where the detection of annihilation products of gravitationally bound
dark matter particles within the core of massive celestial objects like the 
Sun, galaxies, galaxy clusters and dwarf galaxies can provide viable information
about the particle nature of dark matter. These annihilation products include
high energy neutrinos, gamma-rays and charged cosmic rays (electrons,
positrons, protons and antiprotons) \cite{Hooper:2009zm}.
Among these annihilation products gamma-rays and neutrinos
play an important role as they propagate through these celestial objects unperturbed
and thereby directly point towards their sources. 

Recently, several groups have reported 
\cite{Goodenough:2009gk, Hooper:2010mq, Boyarsky:2010dr, Hooper:2011ti,
Abazajian:2012pn, Hooper:2013rwa, Abazajian:2014fta, Daylan:2014rsa,
TheFermi-LAT:2015kwa, Agrawal:2014oha, Calore:2014nla} an excess in gamma-ray flux
in the energy range $1-3$ GeV by analysing the Fermi-LAT publicly available
data \cite{Atwood:2009ez}. These analyses reveal that the observed gamma-ray flux
originates from the inner few degrees around the centre of our Milky Way
galaxy and the nature of this excess gamma-ray spectrum is compatible
with that produced by the annihilation of WIMP dark matter in the Galactic
Centre (GC) region. Although, there are astrophysical explanations
of this anomalous gamma-ray excess in terms of 
unresolved point sources (e.g. millisecond pulsars)
around the GC \cite{Lee:2015fea,
Bartels:2015aea}, in this work we consider a 
dark matter explanation of the GC gamma-ray excess.  
In Ref. \cite{Calore:2014nla}, the authors have
shown that this observed gamma-ray flux
can be well explained by an annihilating self-conjugate dark matter of mass
around $48.7^{+6.4}_{-5.2}$ GeV with an annihilation cross section
${\langle \sigma {\rm v} \rangle}_{b\bar{b}} =
1.75^{+0.28}_{-0.26}\times 10^{-26}$ cm$^3/$s for the $b\bar{b}$ annihilation
channel. In this analysis they have used an NFW \cite{Navarro:1996gj} dark matter
halo profile with $\gamma = 1.26$ and local dark matter density
$\rho_{\odot} = 0.4$ GeV$/$cm$^3$. Moreover, the authors of Ref. \cite{Calore:2014nla}
have considered a region where galactic longitude and latitude
vary in the range $|l|<20^0$ and $2^0<|b|<20^0$ respectively as
the region of interest (ROI) for their analysis.
It is also mentioned in Ref. \cite{Calore:2014nla} that the
uncertainties in the ``astrophysical J factor'', due to
our poor knowledge about DM halo profile parameters,
can change the best fit DM annihilation cross section
${\langle \sigma {\rm v} \rangle}_{b\bar{b}}$
by a multiplicative factor $\mathcal{A}$ which
varies in the range [0.17,5.3]. Various particle dark matter
models explaining the Galactic Centre gamma-ray excess are
available in Refs. \cite{Boucenna:2011hy, Alvares:2012qv, Okada:2013bna,
Alves:2014yha, Berlin:2014tja, Agrawal:2014una, Izaguirre:2014vva,
Cerdeno:2014cda, Ipek:2014gua, Boehm:2014bia, Ko:2014gha, Abdullah:2014lla,
Ghosh:2014pwa, Martin:2014sxa, Wang:2014elb, Basak:2014sza, Detmold:2014qqa,
Arina:2014yna, Okada:2014usa, Ghorbani:2014qpa, Banik:2014eda, Biswas:2014hoa,
Ghorbani:2014gka, Cerdeno:2015ega, Biswas:2015sva, Caron:2015wda,
Borah:2015rla, Banik:2015aya, Dutta:2015ysa, Cuoco:2016jqt}.  

In this work, we consider an extension
of the SM where the gauge sector of SM is enhanced by a local
$\bl$ gauge group where B and L represent
the baryon and lepton numbers respectively. 
In this model we have an extra neutral gauge boson $Z_{BL}$ as
the model Lagrangian possesses an additional local $\bl$ gauge invariance.
In order to construct an anomaly free theory, the model needs three
right handed neutrinos with ${\rm B-L}$ charge equal to $-1$.
Thus, ${\rm B-L}$ extension of the SM is a well motivated beyond Standard Model (BSM)
theory which can explain the origin of tiny neutrino masses through the type-I
seesaw mechanism. Majorana mass terms of these three right handed
neutrinos are generated in a gauge invariant way by introducing a SM gauge singlet
scalar $\phi_H$ having ${\rm B-L}$ charge $+2$. 
The $\bl$ gauge symmetry breaks spontaneously when the scalar field
$\phi_H$ gets a vacuum expectation value $v_{BL}$, thereby
generating mass of the ${\rm B-L}$ gauge boson $Z_{BL}$ and the right handed neutrinos.
The mixing between the neutral components of $\phi_H$ and the SM Higgs
doublet $\phi_h$ produces two physical scalars namely $h_1$ and $h_2$
where $h_1$ is identified as the SM-like Higgs boson with mass around
125.5 GeV. The ${\rm B-L}$ extension of SM \cite{Mohapatra:1980qe,
Georgi:1981pg, Wetterich:1981bx, Lindner:2011it}
has been explored before in the context of dark matter phenomenology
\cite{Basak:2014sza, Okada:2010wd, Okada:2012sg, Basso:2012ti, Basak:2013cga,
Sanchez-Vega:2014rka, Guo:2015lxa, Rodejohann:2015lca, Okada:2016gsh}
and baryogenesis in the early Universe in Refs. \cite{Buchmuller:1992qc,
Buchmuller:1996pa, Dulaney:2010dj}.
In the present work we have introduced a complex scalar
field $\phi_{DM}$ to the $\bl$ extension of SM. This
complex scalar field $\phi_{DM}$ is singlet under the SM gauge group
while it transforms nontrivially under $\bl$ gauge group.  
By choosing proper ${\rm B-L}$ charge, this scalar field
$\phi_{DM}$ can be made stable and hence it can play the role of
a viable dark matter candidate. In this present work, we have
considered the low mass region $40$ GeV to $55$ GeV of DM masses to
explain the Fermi-LAT gamma-ray excess from the Galactic Centre, whereas
the high mass region has been studied in Ref. \cite{Rodejohann:2015lca}.
We have calculated the relic density of $\phi_{DM}$ by
solving Boltzmann equation numerically.
We have found that the gamma-ray flux produced from the annihilation
of $\phi_{DM}$ and $\phi^{\dagger}_{DM}$ can reproduce
the gamma-ray excess as observed by Fermi-LAT from the direction of GC.
Moreover, in this work, we have taken into account all the possible
existing theoretical as well as experimental constraints
obtained from experiments like LHC, LEP, LUX, Planck.  

The paper is arranged in the following way. In Section \ref{model} we describe
the model in detail and discuss the constraints on it from different experiment.
In Section \ref{relic} we calculate the relic density in this model.
In Section \ref{res} we show the variation of the relic density with
different model parameters. In Section \ref{gammaflux} we explain
the Fermi-LAT gamma-ray excess. Finally in Section \ref{summary} we conclude.

\section{Model}
\label{model}
In the present work, we have considered ``pure'' $\bl$ extension of the
Standard Model (SM) of elementary particles where the SM gauge group
$SU(3)_{\rm c} \times SU(2)_{\rm L} \times U(1)_{\rm Y}$
is enhanced by an additional local $\bl$ gauge symmetry where B and L represent
the baryon and lepton numbers, respectively. Therefore, all the SM 
(quarks and leptons) fields transform nontrivially under
this $\bl$ gauge group. Besides the SM fields, we have to
introduce three right handed neutrinos ($N_i,\,i=1$ to 3) such that
the present model becomes anomaly free.
Further, in addition to the usual SM Higgs doublet $\phi_h$, the scalar
sector of the SM is also extended by adding two SM gauge singlet
complex scalar fields, namely $\phi_H$, $\phi_{DM}$
both of which possess nonzero $\bl$ charge. $\bl$ gauge symmetry
breaks spontaneously when the scalar field $\phi_H$ gets a nonzero
vacuum expectation value (VEV) $\vbl$. Consequently,
we have one extra neutral massive gauge field $Z_{BL}$ in the model. Moreover,
after spontaneous breaking of $\bl$ symmetry, the Majorana mass terms for
the three right handed neutrinos can be generated in a gauge invariant
way by choosing a suitable $\bl$ charge $+2$ of the scalar field $\phi_H$.
Also, if the value of the relevant model parameters are such that
the VEV of $\phi_{DM}$ is zero then the complex scalar field $\phi_{DM}$
can be made stable by giving an appropriate ${\rm B-L}$ to it.
Under such circumstances $\phi_{DM}$ can be a viable dark matter candidate.
The $\bl$ charges as well as the SM gauge charges of all the fields present
in the model are given in a tabular form
(see Table \ref{tab1}).    
 
\def\I{i}
\begin{center}
\begin{table}[h!]
\begin{tabular}{||c|c|c|c||}
\hline
\hline
\begin{tabular}{c}
    Gauge\\
    Group\\ 
    \hline
    
    $SU(2)_{L}$\\ 
    \hline
    $U(1)_{Y}$\\ 
    \hline
    $U(1)_{B-L}$\\ 
\end{tabular}
&

\begin{tabular}{c|c|c}
    \multicolumn{3}{c}{Baryon Fields}\\ 
    \hline
    $Q_{L}^{i}=(u_{L}^{i},d_{L}^{i})^{T}$&$u_{R}^{i}$&$d_{R}^{i}$\\ 
    \hline
    $2$&$1$&$1$\\ 
    \hline
    $1/6$&$2/3$&$-1/3$\\ 
    \hline
    $1/3$&$1/3$&$1/3$\\ 
\end{tabular}
&
\begin{tabular}{c|c|c}
    \multicolumn{3}{c}{Lepton Fields}\\
    \hline
    $L_{L}^{i}=(\nu_{L}^{i},e_{L}^{i})^{T}$ & $e_{R}^{i}$ & $N_{R}^{i}$\\
    \hline
    $2$&$1$&$1$\\
    \hline
    $-1/2$&$-1$&$0$\\
    \hline
    $-1$&$-1$&$-1$\\
\end{tabular}
&
\begin{tabular}{c|c|c}
    \multicolumn{3}{c}{Scalar Fields}\\
    \hline
    $\phi_{h}$&$\phi_{H}$&$\phi_{DM}$\\
    \hline
    $2$&$1$&$1$\\
    \hline
    $1/2$&$0$&$0$\\
    \hline
    $0$&$2$&$n_{BL}$\\
\end{tabular}\\
\hline
\hline
\end{tabular}
\caption{Particle content and their corresponding
charges under various symmetry groups.}
\label{tab1}
\end{table}
\end{center} 

The Lagrangian of the present model including the
SM Lagrangian $\mathcal{L}_{SM}$ is as follows
\begin{eqnarray}
\mathcal{L}&=&\mathcal{L}_{SM} + \mathcal{L}_{DM} +
(D_{\mu}\phi_{H})^{\dagger} (D^{\mu}\phi_{H})
-\frac{1}{4} {F_{BL}}_{\mu \nu} {F_{BL}}^{\mu \nu} + 
\frac{i}{2}\bar{N_i}\gamma^{\mu}D_{\mu} N_{i}
-V(\phi_{h},\phi_{H})\nn \\
&&-\sum_{i=1}^3 \frac{1}{2} \lambda_{N_{i}}
\phi_{H} \bar{N^{c}_{i}} N_{i} -\sum_{i,\,j=1}^3 y_{ij} \bar{L_{i}}
\tilde {\phi_{h}} N_{j} +h.c.\,,
\label{lag}
\end{eqnarray}
with $\tilde{\phi_h} = i \sigma_2 \phi^*_h$, while $\mathcal{L}_{DM}$
represents the dark sector Lagrangian whose expression is given by
\begin{eqnarray}
\mathcal{L}_{DM} &=& (D^{\mu}\phi_{DM})^\dagger (D_{\mu}\phi_{DM})
- \mu_{DM}^{2} \phi_{DM}^{\dagger} \phi_{DM} 
-\lambda_{Dh}(\phi_{DM}^{\dagger} \phi_{DM})
(\phi_{h}^{\dagger} \phi_{h}) \nn \\ && 
-\lambda_{DH}(\phi_{DM}^{\dagger} \phi_{DM})(\phi_{H}^{\dagger} \phi_{H})
-\lambda_{DM} (\phi_{DM}^{\dagger} \phi_{DM})^{2} \,,
\label{ldm}
\end{eqnarray}
and the self interactions of $\phi_H$ and its mutual interaction
with the SM Higgs doublet $\phi_h$ are described
by $V(\phi_h, \phi_H)$ which can be written as
\begin{eqnarray}
V(\phi_h, \phi_H) = \mu_{H}^{2} \phi_{H}^{\dagger} \phi_{H} 
+ \lambda_{H} (\phi_{H}^{\dagger} \phi_{H})^{2}
+ \lambda_{hH}(\phi_{h}^{\dagger} \phi_{h}) (\phi_{H}^{\dagger} \phi_{H}) \,.
\label{int}
\end{eqnarray}
In Eq. (\ref{lag}), ${F_{BL}}_{\mu \nu}=\partial_{\mu} {Z_{BL}}_{\nu} -
\partial_{\nu} {Z_{BL}}_{\mu}$ is the field strength tensor of the $\bl$
gauge field $Z_{BL}$. Covariant derivative appearing in Eqs. (\ref{lag}, \ref{ldm})
is defined as
\begin{eqnarray}
D_\mu \psi = (\partial_{\mu} + i\,g_{BL}\,Q_{BL}(\psi)\,{Z_{BL}}_{\mu})\,\psi \,,
\end{eqnarray} 
where $\psi = \phi_H$, $\phi_{DM}$, $N_i$ and $Q_{BL}(\psi)$ is the
corresponding $\bl$ gauge charge which is given in Table \ref{tab1}.
In general, the Majorana mass matrix for the three right handed neutrinos,
obtained after spontaneous breaking of ${\rm B-L}$ symmetry,
will contain {\it off diagonal} terms. However these off diagonal terms
can be easily removed by changing the basis and therefore, we have considered the diagonal
Majorana mass matrix for the three right handed neutrinos (or the right handed
neutrinos $N_i$'s are in mass basis). 

After spontaneous breaking of $\bl$ symmetry the scalar fields
$\phi_{h}$ and $\phi_{H}$ in unitary gauge take the following form
\begin{eqnarray}
\phi_{h}=
\begin{pmatrix}
0 \\
\dfrac{v+H}{\sqrt{2}}
\end{pmatrix}
\,\,\,\,\,\,\,\,\,
\phi_{H}=
\begin{pmatrix}
\dfrac{v_{BL}+H_{BL}}{\sqrt{2}}
\end{pmatrix}\,\,,
\label{phih}
\end{eqnarray}
where $v = 246\,\,{\rm GeV}$ is the VEV of $\phi_h$, which breaks the
electroweak symmetry to a $U(1)$ symmetry ($U(1)_{em}$). On the other hand
the VEV of $\phi_H$, $v_{BL}$, is responsible for the breaking of ${\rm B-L}$
gauge symmetry of the Lagrangian and thereby generates masses for the three
right handed neutrinos as well as the gauge boson $Z_{BL}$, 
\begin{eqnarray}
M_{N_i} &=& \dfrac{\lambda_{N_i}}{\sqrt{2}} v_{BL} \,, \nn \\
M_{Z_{BL}} &=& 2\,g_{BL}\,v_{BL}\,.
\end{eqnarray}
In Eq. (\ref{phih}) $H$ and $H_{BL}$ are two neutral scalar
fields of $\phi_h$ and $\phi_H$ respectively. There is also mixing between $H$ and $H_{BL}$
through the term $\lambda_{hH}$ (see Eq. (\ref{int})).
As a result, the mass matrix of $H$ and $H_{BL}$ contains
off diagonal elements which are proportional
to $\lambda_{hH}$, $v$ and $V_{BL}$. Hence, $H$ and $H_{BL}$ are
not representing any physical field. The scalar mass matrix with respect
to the basis $(H,\,\,H_{BL})$ is given by
\begin{eqnarray}
\mathcal{M}^2_{scalar} = \left(\begin{array}{cc}
2\lambda_h v^2 ~~&~~ \lambda_{hH}\,v_{BL}\,v \\
~~&~~\\
\lambda_{hH}\,v_{BL}\,v ~~&~~ 2 \lambda_H v^2_{BL}
\end{array}\right) \,\,.
\label{mass-matrix}
\end{eqnarray}
In order to obtain the physical states we have to diagonalise the
real \footnote{In present model we have taken all the coupling
constants and VEVs as real.} symmetric matrix $\mathcal{M}^2_{scalar}$
(Eq. (\ref{mass-matrix}))
by an orthogonal matrix. The physical fields or the mass eigenstates
which are linearly related to $H$ and $H_{BL}$, can be obtained
through the following relations 
\begin{eqnarray}
h_{1}&=& H \cos \alpha + H_{BL} \sin \alpha \,, \nn \\
h_{2}&=& - H \sin \alpha + H_{BL} \cos \alpha\,,
\end{eqnarray}
where the scalar field $h_1$ is identified as the SM like Higgs
boson and $h_2$ is the extra Higgs boson in the model, while $\alpha$
is the mixing angle between $H$ and $H_{BL}$ given as
\begin{eqnarray}
\tan 2\alpha = \dfrac{\lambda_{hH}\,v_{BL}\,v}
{\lambda_h v^2 - \lambda_H v^2_{BL}}\,.
\end{eqnarray}
We will see later that from LHC results, the allowed values of
the mixing angle $\alpha$ are extremely small. The expressions
of masses of the three physical scalar fields $h_1$, $h_2$ and $\phi_{DM}$
are  
\begin{eqnarray}
M^2_{h_1} &=& \lambda_h v^2 + \lambda_H v_{BL}^2 + 
\sqrt{(\lambda_h v^2 - \lambda_H v_{BL}^2)^2 + (\lambda_{hH}\,v\,v_{BL})^2}
\ ,\nn \\
M^2_{h_2} &=& \lambda_h v^2 + \lambda_H v^2_{BL} - 
\sqrt{(\lambda_h v^2 - \lambda_H v_{BL}^2)^2 + (\lambda_{hH}\,v\,v_{BL})^2}\,,
\nn \\
M^2_{DM} &=& \mu^2_{DM} + \dfrac{\lambda_{Dh} v^2}{2} +
\dfrac{\lambda_{DH} v^2_{BL}}{2}\,.
\end{eqnarray}
Since $h_1$ is the SM like Higgs boson therefore we have taken $M_{h_1} = 125.5$ GeV. 

In this model, besides the SM parameters, we have twelve unknown independent
parameters, namely the masses of $h_2$, $\phi_{DM}$, $Z_{BL}$, $N_i$,
$\bl$ gauge coupling $g_{BL}$, ${\rm B-L}$ charge ($n_{BL}$)
of dark matter ($\phi_{DM}$),
scalar mixing angle $\alpha$ and three quartic couplings $\lambda_{DH}$, $\lambda_{Dh}$,
$\lambda_{DM}$. In terms of these independent parameters, the couplings appearing in
the Lagrangian (Eqs. (\ref{lag}-\ref{int})) can be expressed as
\begin{eqnarray}
\lambda_{H} &=& \dfrac{M_{h_{1}}^{2} + M_{h_{2}}^{2} +
(M_{h_{2}}^{2} - M_{h_{1}}^{2})\cos 2 \alpha}{4\,v_{BL}^{2}}\,,\nn \\
\lambda_{h}&=& \dfrac{M_{h_{1}}^{2} + M_{h_{2}}^{2} +
(M_{h_{1}}^{2} - M_{h_{2}}^{2})\cos 2 \alpha}{4\,v^{2}}\,,\nn\\
\lambda_{hH} &=& \dfrac{(M_{h{1}}^{2}-M_{h_{2}}^{2})
\cos \alpha \sin \alpha}{v\,v_{BL}}\,,\nn\\
\mu_{\phi_{h}}^{2}&=& - \dfrac{(M_{h_1}^{2}+M_{h_2}^{2})v
+(M_{h_1}^{2}-M_{h_2}^{2})(v \cos 2 \alpha + v_{BL}
\sin 2 \alpha)}{4\,v} \,,\nn\\
\mu_{\phi_{H}}^{2}&=& \dfrac{-(M_{h_1}^{2}+M_{h_2}^{2})v_{BL}
+ (M_{h_1}^{2}-M_{h_2}^{2})(v_{BL}\cos 2 \alpha - v
\sin 2 \alpha)}{4\,v_{BL}}\,,
\end{eqnarray}
where $\mu^2_{\phi_h}$ and $\mu^2_{\phi_H}$ are the quadratic self coupling of
the SM Higgs doublet $\phi_h$ and the extra Higgs singlet $\phi_H$ respectively.
Moreover, the model parameters are subjected to satisfy certain conditions 
arising from theoretical constraints as well as relevant experimental results. 
These constraints are briefly discussed
below.
\begin{itemize}
\item {\bf Vacuum Stability:}
In our model we choose the ground state ($\phi_h$, $\phi_H$, $\phi_{DM}$)
= ($v$, $v_{BL}$, 0). This requires the following constrains on the
quadratic self couplings of the scalar fields,  
\begin{eqnarray}
\mu^2_{\phi_h} < 0, \,\,\,\,\,\,\mu^2_{\phi_H} < 0\,\,\,\, {\rm and}\,\,\,\,
\mu^2_{DM} > 0\,. 
\end{eqnarray} 
Also in order to obtained a stable ground state ({\it vacuum}), the
quartic couplings, appearing in the Lagrangian, need to satisfy the
following conditions
\begin{eqnarray}
&&\lambda_h \geq 0, \lambda_H \geq 0, \lambda_{DM} \geq 0,\nonumber \\
&&\lambda_{hH} \geq - 2\sqrt{\lambda_h\,\lambda_H},\nonumber \\
&&\lambda_{Dh} \geq - 2\sqrt{\lambda_h\,\lambda_{DM}},\nonumber \\
&&\lambda_{DH} \geq - 2\sqrt{\lambda_H\,\lambda_{DM}},\nonumber \\
&&\sqrt{\lambda_{hH}+2\sqrt{\lambda_h\,\lambda_H}}\sqrt{\lambda_{Dh}+
2\sqrt{\lambda_h\,\lambda_{DM}}}
\sqrt{\lambda_{DH}+2\sqrt{\lambda_H\,\lambda_{DM}}} \nonumber \\ 
&&+ 2\,\sqrt{\lambda_h \lambda_H \lambda_{DM}} + \lambda_{hH} \sqrt{\lambda_{DM}}
+ \lambda_{Dh} \sqrt{\lambda_H} + \lambda_{DH} \sqrt{\lambda_h} \geq 0 \,\,\,\,.
\end{eqnarray}
\item {\bf Planck Limit:} The relic density $\Omega_{DM} h^2$ of the dark matter
particle $\phi_{DM}$ at the present epoch should lie within the range
reported by the satellite borne experiment Planck \cite{Ade:2015xua}, which is
\begin{eqnarray}
0.1172\leq \Omega_{DM} h^2 \leq 0.1226\,\,\,\,\,\,{\rm at}\,\,68\%\,\,{\rm C.L.}
\end{eqnarray}
\item {\bf Stability of dark matter:}
We give a $\bl$ charge to the dark matter
candidate ($\phi_{DM}$) in such a way so that all possible decay terms are
forbidden by the invariance of $\bl$ gauge symmetry which therefore ensures the
stability of $\phi_{DM}$. In general, the possible decay terms of $\phi_{DM}$
are like $\phi_{DM} \phi_{h}^{p} \phi_{H}^{q}$ (where $p+q \leq 3$ and
$p$, $q$ are integer can vary from $0$ to $3$) and
$\phi_{DM} \bar{f^{\prime}} f$, where $f$ is $N_i$
and $f^{\prime}=N^c_i$ \footnote{Since $\phi_{DM}$ is singlet
under SM gauge group therefore a term like $\phi_{DM} {f} \bar{f}$ with $f$
being any Standard Model fermion is forbidden.}.
From Table \ref{tab1} one can see that the ${\rm B-L}$ charges
of $\phi_h$ and $\phi_H$ are 0 and +2 respectively. Therefore if we take
$n_{BL} \neq - 2\,q$ then we can not write the term
$\phi_{DM} \phi_{h}^{p} \phi_{H}^{q}$, as it will violate the $\bl$
gauge symmetry. In addition, in our case, we have varied dark matter
mass from $40$ GeV to $55$ GeV and $M_{DM} < M_{h_1},\,M_{h_2}$
\footnote{which is required to explain Fermi-LAT gamma excess
\cite{Calore:2014nla}, see section \ref{gammaflux} for more detailed discussion.}
as a result any decay modes of $\phi_{DM}$ to these scalar bosons
are kinematically forbidden.
Moreover, due to the presence of $\phi_{DM} \bar{N^c_i} N_i$ term,
the dark matter candidate can also decay into two Majorana type
right handed neutrinos in the final state, if the kinematical
condition ($M_{DM} > 2 M_{N_i}$, $i=$ 1 to 3) is satisfied, which
can also destroy its stability. To get rid of this decay term
we can not choose $n_{BL} = + 2$ as the combination $\bar{N_i^c} N_i$
has ${\rm B-L}$ charge $-2$. Therefore, in order to avoid all the above
mentioned decay terms (due to renormalizability of the Lagrangian
we have considered operators only upto dimension 4)
we need $n_{BL}\neq \pm 2\,q$ where $q$ is any
integer between 0 and 3.  
\item {\bf LEP bound:}
Since the SM fermions are charged under the gauge group $\bl$, therefore
LHC should find some footprint of the ${\rm B-L}$ gauge boson $Z_{BL}$
as it can directly interact with all the SM fermions.
The nondetection of any signature
of $Z_{BL}$ puts a severe constraint on its mass ($M_{Z_{BL}}$)
and ${\rm B-L}$ gauge coupling ($g_{BL}$). From LEP experiment the
ratio $\frac{M_{Z_{BL}}}{g_{BL}}$ is bounded from below by the following
condition \cite{Carena:2004xs,Cacciapaglia:2006pk} \footnote{Recent
bounds on the mass and gauge coupling of $Z_{\rm BL}$
from ATLAS collaboration are given in Ref. \cite{Aad:2014cka}.}
\begin{eqnarray}
\frac{M_{Z_{BL}}}{g_{BL}} \ga 6-7 \,\, {\rm TeV}\,.
\end{eqnarray}
\begin{figure}[h!]
\centering
\includegraphics[angle=0,height=5cm,width=16cm]{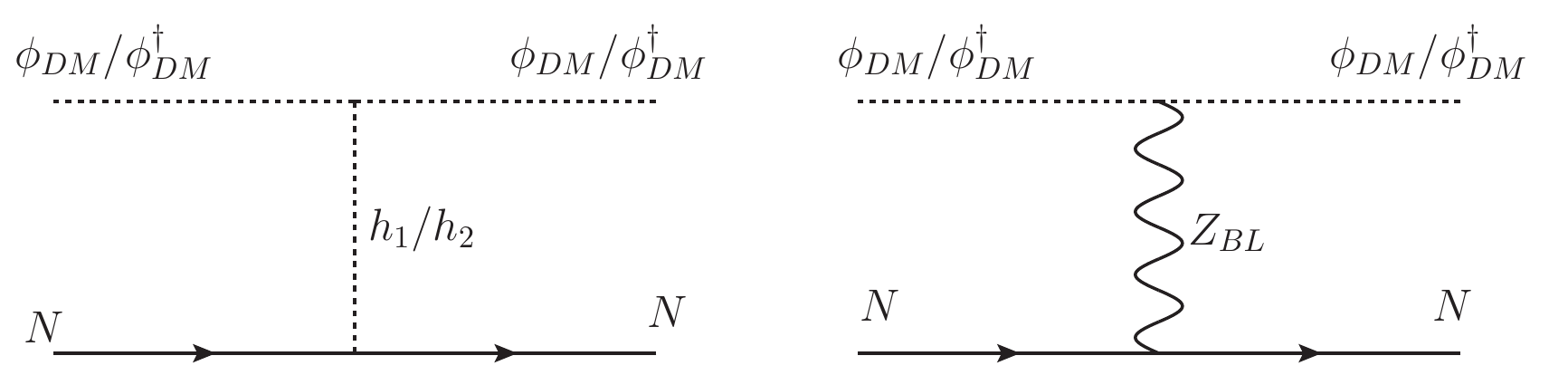}
\caption{Feynmann diagrams for spin independent scattering cross section of
dark matter particle/antiparticle with nucleon ($N$)
through both Higgses ($h_1,h_2$) and gauge boson $Z_{BL}$. }
\label{feyn2}
\end{figure}
\item {\bf LUX limit:}
In this model, the complex scalar field $\phi_{DM}$
is our dark matter candidate. Therefore, both $\phi_{DM}$ and its antiparticle can
elastically scatter off the detector nuclei through the exchange of neutral scalars
$h_1$, $h_2$ and $\bl$ gauge boson $Z_{BL}$. Moreover, due to the
presence of vector boson ($Z_{BL}$) mediator, the elastic scattering
cross sections for the dark matter and its antiparticle are different. 
If we take the number densities of the dark matter and its antiparticle to be equal 
at the present epoch (which is true if the species has negligible
chemical potential \cite{Gondolo:1990dk}), then we have to multiply the elastic scattering
cross sections of dark matter and its antiparticle by a factor $1/2$
while comparing these scattering cross sections, obtained from the present
model, with the experimental upper limits reported by the direct detection
experiment LUX \cite{Akerib:2013tjd, Akerib:2015rjg}.
The reason behind this is the exclusion regions
in $\sigma_{SI}-M_{DM}$ plane reported by different dark matter direct detection
experiments are computed assuming the existence of only one type of
dark matter particle (and also self-conjugate) in the Universe. Although 
our model too has only one kind of dark matter candidate, however,
it has a different antiparticle and they do not possess equal interaction strengths
with the detector nuclei. Feynman diagrams for the elastic scattering of
both $\phi_{DM}$ and $\phi^{\dagger}_{DM}$ with the nucleon ($N$) are
shown in Fig.\,\,\ref{feyn2}. These processes are mediated through
the exchange of $h_1$, $h_2$ and $Z_{BL}$. The expressions of spin
independent scattering cross sections off the nucleon ($N$)
for both $\phi_{DM}$ and $\phi^{\dagger}_{DM}$ are given by

\begin{eqnarray}
\sigma_{\phi_{DM}\,({\phi^{\dagger}_{DM}})}
= \dfrac{\mu^2}{4\pi}\Bigg[\frac{M_N\,f_N\,\cos \alpha}{M_{DM}\,v}
\Bigg(\frac{\tan \alpha\,g_{\phi_{DM}\phi^{\dagger}_{DM} h_2}}{M^2_{h_2}}
-\frac{g_{\phi_{DM}\phi^{\dagger}_{DM} h_1}}{M^2_{h_1}}
\Bigg) -(+)
\frac{2\,n_{BL}\,g^2_{BL}\,f_{Z_{BL}}}{3\,M^2_{Z_{BL}}}
\Bigg]^2,\nonumber\\
\label{sigmasi}
\end{eqnarray}
where $g_{\phi_{DM}\phi^{\dagger}_{DM} h_i}$ is the vertex factor
for a vertex involving fields
$\phi_{DM}\,\phi^{\dagger}_{DM}\,h_i$ ($i=1$, 2) and
its expression is given in Table \ref{tab1}. The reduced mass between nucleon $N$
(proton or neutron) and DM particle is denoted by $\mu$. Moreover, the nuclear
form factor for the scalar mediated processes is $f_N \sim 0.3$
\cite{Cline:2013gha} while that for $Z_{BL}$ mediated
diagram is $f_{Z_{BL}} = 3.0$ \footnote{ $N\,\bar{N}\,Z_{\rm BL}$
coupling $g_{{}_{N\,{\bar{N}}\,Z_{\rm BL}}}=
\sum_{q = u\, d} f^N_{V_q} \times g_{q\,\bar{q}\,Z_{\rm BL}}$
\cite{Belanger:2008sj} with $g_{q\,\bar{q}\,Z_{\rm BL}} = -\frac{g_{\rm BL}\,
\gamma^{\mu}}{3}$
is the coupling for the vertex containing fields
$q \, \bar{q}\, Z_{\rm BL}$ (see Table \ref{tab2}).
Now for proton $p$ (neutron $n$)
$f^p_{V_u} = 2,\,f^p_{V_d} = 1$ ($f^n_{V_u} = 1,\,f^n_{V_d} = 2$)
\cite{Belanger:2008sj}. Therefore, for both the nucleon
$N$ ($n$ and $p$) the coupling $g_{{}_{N\,{\bar{N}}\,Z_{\rm BL}}}=
f_{Z_{\rm BL}}\,\times g_{q\,\bar{q}\,Z_{\rm BL}}$ with $f_{Z_{\rm BL}}=
\sum_{q = u\, d} f^N_{V_q} = 3$. Thus in this model form factors
of proton and neutron are same for $Z_{\rm BL}$ mediated
diagram.}.
From the expression of spin independent scattering cross section
it is seen that although, the elastic scattering cross sections of
$\phi_{DM}$ and $\phi^{\dagger}_{DM}$ with $N$ are identical when
the scattering processes are mediated through the scalar bosons only,
however, if we include the $Z_{BL}$ mediated diagram then
the elastic scattering cross sections for both $\phi_{DM}$
and $\phi^{\dagger}_{DM}$ become different from each other.
It is due to the fact that the momentum dependent vertex
factors for the vertices $\phi_{DM} \phi_{DM} Z_{BL}$
and $\phi^{\dagger}_{DM} \phi^{\dagger}_{DM} Z_{BL}$
\footnote{see the expression of $g_{\phi_{DM} \phi^{\dagger}_{DM}
Z_{BL}}$ in Table \ref{tab2}} are
differ by a -ve sign (due to the change in sign of momentum while
go from particle to anti particle scenario) from each other
which results in a difference between $\sigma_{\phi_{DM}}$
and $\sigma^{\dagger}_{\phi_{DM}}$ arising from the interaction
terms between $Z_{BL}$ and scalar bosons mediated diagrams. 
If $\sigma^{exp}_{SI}$ represents the upper limit
of the spin independent scattering cross section reported by the
LUX experiment for a particular dark matter mass then for a
viable dark matter model both $\sigma_{\phi_{DM}}$ and
$\sigma_{\phi^{\dagger}_{DM}}$ must satisfy the following condition
\begin{eqnarray}
\sigma_{\phi_{DM}} + \sigma_{\phi^{\dagger}_{DM}} < 2\sigma^{exp}_{SI}\,,
\end{eqnarray}
\item {\bf LHC constraints:}
\begin{itemize}
\item{\bf Signal Strength of SM-like Higgs:}
The signal strength of $h_1$ for a particular
decay channel $h_1\rightarrow X \bar{X}$ ($X$ is any SM
particle such as gauge boson, quark or lepton)
is defined as
\begin{eqnarray}
R_{X\bar{X}} = \dfrac{\sigma\,BR(h_1\rightarrow X \bar{X})}
{\,\,\,\,\,\,[\sigma\,BR(h \rightarrow X \bar{X})]_{SM}} \,,
\label{rxx}
\end{eqnarray}
where $\sigma$ and $BR(h_1 \rightarrow X \bar{X})$ are the
production cross section of $h_1$ and its branching ratio
for $X \bar{X}$ decay channel. In the denominator of the
above equation $[\sigma\,BR(h \rightarrow X \bar{X})]_{SM}$
represent the same quantities for the SM Higgs boson ($h$).
If the neutral boson $h_1$ is similar to the SM Higgs boson
then according to LHC result the signal strength ratio
$R_{X\bar{X}}$ should be $>0.8$ \footnote{We have considered
the central value of the combined signal strength of the
SM Higgs boson reported by the CMS collaboration \cite{Agashe:2014kda}.}
\cite{Agashe:2014kda}. We will see later,
in Fig. \ref{1a} (Section \ref{res}) that the above condition will impose
severe constrain on the allowed values of scalar mixing angle $\alpha$.

\item {\bf Invisible decay width of Higgs boson:} In the present model, the
SM like Higgs boson $h_1$ can decay into a pair of $\phi_{DM}$ and
$\phi^{\dagger}_{DM}$ if the kinematical condition $M_{h_1}\geq 2 M_{DM}$
is satisfied. Such decay channel is known as the invisible decay
model of $h_1$. The expression of partial decay width of $h_1$
into $\phi_{DM} \phi^{\dagger}_{DM}$ final state is
\begin{eqnarray}
\Gamma_{h_{{}_1} \rightarrow \phi_{{}_{{}_{DM}}}
\phi^{\dagger}_{DM}} = 
\dfrac{g^2_{h_{{}_1} \phi_{{}_{{}_{DM}}}
\phi^{\dagger}_{DM}}}
{ 16 \pi \,M_{h_{{}_1}}}
\sqrt{1-\dfrac{4 M^2_{DM}}{M^2_{h_{{}_1}}}} \,,
\label{h1inv}
\end{eqnarray}
where $g_{h_{{}_1} \phi_{{}_{{}_{DM}}} \phi^{\dagger}_{DM}}$ is the vertex
factor for the vertex involving $h_{1} \phi_{{}_{{}_{DM}}} \phi^{\dagger}_{DM}$.
Throughout this work we have considered
the partial width of this invisible decay channel of $h_1$
to be less than 20\% \cite{Bechtle:2014ewa, Belanger:2013kya}
of its total decay width.
\end{itemize}
\item{\bf Fermi-LAT gamma excess from Galactic Centre:} In order to
explain the Fermi-LAT observed gamma-ray excess from the Galactic Centre
using a self-conjugate annihilating dark matter, one needs a dark matter
particle of mass $48.7^{+6.4}_{-5.2}$ GeV \cite{Calore:2014nla}. 
If we assume an NFW halo profile with $\gamma=1.26$,
$\rho_{\odot} =0.4$ GeV/cm$^3$, $r_{\odot}=8.5$ kpc and
$r_s = 20$ kpc then the annihilation cross section of dark matter
particle for the $b\bar{b}$ annihilation channel should lie
in the range $\langle {\sigma {\rm v}} \rangle_{b \bar{b}} \sim
1.75^{+0.28}_{-0.26} \times 10^{-26}$ cm$^3/$s \cite{Calore:2014nla}. However,
if we take into account the uncertainties of DM halo profile
parameters (mentioned above) then the quantity
$\langle {\sigma {\rm v}} \rangle_{b \bar{b}}$ can vary
in the range $\mathcal{A} \times 1.75^{+0.28}_{-0.26} \times 10^{-26}$
cm$^3$/s with $\mathcal{A}=[0.17,5.3]$ \cite{Calore:2014nla}. We will discuss
about the Fermi-LAT gamma-ray excess elaborately in Section \ref{gammaflux}
where we will see that the required value of
$\langle {\sigma {\rm v}} \rangle_{b \bar{b}}$ for a non-self-conjugate DM
(which is true for the present model) is different from a dark
matter candidate whose particles and antiparticles are same.
\end{itemize}
\begin{figure}[h!]
\centering
\includegraphics[angle=0,height=5cm,width=16cm]{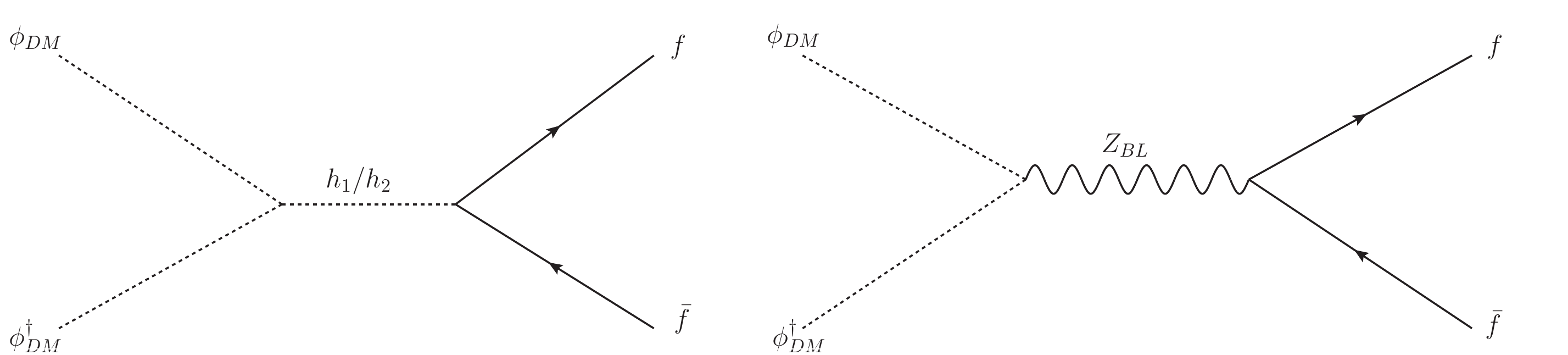}
\caption{Feynman diagrams for dark matter annihilation through both scalar
bosons ($h_1,h_2$) and gauge boson $Z_{BL}$. }
\label{feyn1}
\end{figure}
\section{Relic Density}
\label{relic}
The evolution of total number density ($n$) of both $\phi_{DM}$
and $\phi^{\dagger}_{DM}$ is governed by the Boltzmann equation
which is given by \cite{Gondolo:1990dk} 
\begin{eqnarray}
\frac{dn}{dt} + 3 n{\rm H} = 
-\frac{1}{2}\langle{\sigma {\rm{v}}}\rangle 
\left(n^2 -(n^{eq})^2\right) \,,
\label{be}
\end{eqnarray}
where ${\rm H}$ is the Hubble parameter and $n^{eq}$ is the
equilibrium number density of both $\phi_{DM}$ and $\phi^{\dagger}_{DM}$.
$\sigma$ is the annihilation cross section for the channel
$\phi_{DM} \phi^{\dagger}_{DM} \rightarrow f \bar{f}$, where $f$ is
any SM fermion except top quark \footnote{
In order to explain Fermi-LAT $\gamma$-ray excess we need
$M_{DM}$ in the range $48.7^{+6.4}_{-5.2}$ GeV \cite{Calore:2014nla}
and thus other annihilation channels of $\phi_{DM}$
($\phi_{DM} \phi^{\dagger}_{DM} \rightarrow W^+ W^-$,
$ZZ$, $Z_{BL} Z_{BL}$, $t\bar{t}$, $h_1 h_1$ etc.)
are not kinematically allowed.}. Tree level Feynman diagrams
for the process $\phi_{DM} \phi^{\dagger}_{DM} \rightarrow f \bar{f}$
mediated through the exchange of $h_1$, $h_2$ and $Z_{BL}$
are given in Fig. {\ref{feyn1}}. The expression of
$\sigma$ is as follows,
\begin{eqnarray}
\sigma &=& \frac{3}{8 \pi s}
\sqrt{\dfrac{s-4m_{f}^{2}}{s-4 M_{DM}^{2}}}\Bigg\{A^2\,(s-4m_{f}^{2}) 
\left|\frac{g_{h_{1} \phi_{{{}_{DM}}} \phi^{\dagger}_{DM}}}
{(s-M_{h_{1}}^{2}) + \I \Gamma_{h_{1}} M_{h_{1}}} -
\frac{\tan \alpha\,\,
g_{h_{2} \phi_{{{}_{DM}}} \phi^{\dagger}_{DM}}}{(s-M_{h_{2}}^{2}) + \I
\Gamma_{h_{2}} M_{h_{2}}} \right|^{2}  \nn \\
&&
+ \dfrac{2}{9} \dfrac{g_{BL}^{4} n_{BL}^{2}}{(s- M_{Z_{BL}})^{2}
+ (\Gamma_{Z_{BL}} M_{Z_{BL}})^2}
(s-4 M_{DM}^{2})(s+ 2 m_{f}^{2})\Bigg\},
\label{sigmaff}
\end{eqnarray}  
where $\Gamma_i$ is the total decay width of the particle
$i$ ($i=h_1$, $h_2$, $Z_{BL}$), $m_f$ is the mass of the
SM fermion $f$ and $\sqrt{s}$ is centre of mass energy.
$g_{i\,\phi_{{{}_{DM}}} \phi^{\dagger}_{DM}}$ is the
vertex factor for the vertex involving the fields
$i\,\phi_{{{}_{DM}}} \phi^{\dagger}_{DM}$ ($i=h_1$, $h_2$)
and its expression is given in Table \ref{tab2}. Moreover the quantity
$A=\dfrac{m_{f}}{v} \cos \alpha$, with $\alpha$
is the scalar mixing angle, is the coupling
for the vertex $f\,\bar{f}\,h_1$ (see Table \ref{tab2}).
In Eq. (\ref{be}), $\langle{\sigma {\rm{v}}}\rangle$
represents the thermal average of the product
between annihilation cross section $\sigma$ and the relative
velocity ${\rm v}$ of the annihilating particles. Extra $1/2$
factor appearing before $\langle{\sigma {\rm{v}}}\rangle$
is due to non-self-conjugate nature of $\phi_{DM}$ \cite{Gondolo:1990dk}.
Thermal averaged annihilation cross section $\langle{\sigma {\rm{v}}}\rangle$ 
can be defined in terms of annihilation cross section $\sigma$ and modified
Bessel functions ($K_1,\,K_2$) as \cite{Gondolo:1990dk}
\begin{eqnarray}
\left\langle \sigma {\rm v}\right\rangle 
&=& \frac{1}{8 M_{DM}^4 T K_2^2\left(\frac{M_{DM}}{T}\right)}
\int_{4M_{DM}^2}^\infty \sigma\,(s-4M_{DM}^2)\,\sqrt{s}\,K_1
\left(\frac{\sqrt{s}}{T}\right)\,ds \,\, ,
\label{thermal-ave}
\end{eqnarray}
where $T$ is the temperature of the Universe.
Now, we define two dimensionless variables $Y$ and $x$ as follows
\begin{eqnarray*}
Y = \dfrac{n}{\rm s}\,,\,\,\,\,\,\,\,x=\dfrac{M_{DM}}{T}
\end{eqnarray*}
with ${\rm s}$ is the entropy density of the Universe while
$Y$ is called the total comoving number density of both $\phi_{DM}$
and $\phi^{\dagger}_{DM}$.
In terms of these two variables the Boltzmann equation,
given in Eq. (\ref{be}), can be written as
\begin{eqnarray}
\frac{dY}{dx} =
-\left(\frac{45G}{\pi}\right)^{-\frac{1}{2}}
\frac{M_{DM}\,\sqrt{g_\star}}{x^2}\,
\frac{1}{2} \langle{\sigma {\rm{v}}}\rangle
\left(Y^2-(Y^{eq})^2\right)\,,
\label{be1}
\end{eqnarray}
where $G$ is the Newton's gravitational constant and $g_{\star}$
is a function of effective degrees of freedom related to both
energy and entropy densities of the Universe \cite{Gondolo:1990dk}. Now, one can
find the value of the total comoving number density ($Y$) at the present temperature
($T_0\sim 10^{-13}$ GeV) by solving the first order differential
equation. The estimated value of $Y(T_0)$ is then used to compute
the total relic abundance of both $\phi_{DM}$ and $\phi^{\dagger}_{DM}$
at the present epoch through the following equation \cite{Edsjo:1997bg}
\begin{eqnarray}
\Omega_{DM} h^2 = 2.755\times 10^8 \left(\frac{M_{DM}}{\rm GeV}\right) Y(T_0)\,.
\end{eqnarray}
\begin{table}[h!]
\begin{center}
\vskip 0.5cm
\begin{tabular} {||c||c||}
\hline
\hline
Vertex & Vertex Factor\\
$a\,b\,c$ & $g_{abc}$\\
\hline
$q\,\bar{q}\,h_1$ & $-\dfrac{M_{q}}{v} \cos \alpha$\\
\hline
$q\,\bar{q}\,h_2$ & $ \dfrac{M_{q}}{v} \sin \alpha$\\
\hline
$q\,\bar{q}\,Z_{BL}$ & $-\dfrac{g_{BL}}{3}\,\gamma^\mu$\\
\hline
$l\,\bar{l}\,h_1$ & $-\dfrac{M_{l}}{v} \cos \alpha$\\
\hline
$l\,\bar{l}\,h_2$ & $ \dfrac{M_{l}}{v} \sin \alpha$\\
\hline
$l\,\bar{l}\,Z_{BL}$ & ${g_{BL}}\,\gamma^\mu$\\
\hline
$\phi_{DM}\,\phi^{\dagger}_{DM}\,h_1$ &$-\dfrac{1}{2\,g_{BL}}
(2\,g_{BL}\,v\,\lambda_{Dh} \cos \alpha + M_{Z_{BL}} \lambda_{DH} \sin \alpha)$ \\
\hline
$\phi_{DM}\,\phi^{\dagger}_{DM}\,h_2$ & $\dfrac{1}{2\,g_{BL}}
(2\,g_{BL}\,v\,\lambda_{Dh} \sin \alpha - M_{Z_{BL}} \lambda_{DH} \cos \alpha)$\\
\hline
$\phi_{DM}\,\phi^{\dagger}_{DM}\,Z_{BL}$ & $n_{BL}\,g_{BL}\,(p_{2}-p_{1})^{\mu}$\\
\hline
$\phi_{DM}\,\phi^{\dagger}_{DM}\,h_1\,h_1$ & $-(\lambda_{Dh}\,
\cos^2 \alpha  + \lambda_{DH}\,\sin^2 \alpha)$\\
\hline
$\phi_{DM}\,\phi^{\dagger}_{DM}\,h_2\,h_2$ & $-(\lambda_{Dh}\,
\sin^2 \alpha  + \lambda_{DH}\,\cos^2 \alpha)$\\
\hline
$\phi_{DM}\,\phi^{\dagger}_{DM}\,h_1\,h_2$ &
$\sin \alpha \cos \alpha (\lambda_{Dh} - \lambda_{DH})$\\
\hline
$\phi_{DM}\,\phi^{\dagger}_{DM}\,Z_{BL}\,Z_{BL}$ & 
$2\,g^2_{BL} n^2_{BL}$\\
\hline
$\phi_{DM}\,\phi^{\dagger}_{DM} \phi_{DM}\,\phi^{\dagger}_{DM}$ &
$-4\,\lambda_{DM}$\\
\hline
\hline
\end{tabular}
\end{center}
\caption{All possible vertex factors related to dark matter
annihilation for the present model.}
\label{tab2}
\end{table}
\section{Results}
\label{res}
In this section we have shown how the relic density of DM varies with
various model parameters namely $\alpha$, $g_{BL}$, $n_{BL}$,
$M_{h_2}$, $M_{DM}$, $M_{Z_{BL}}$, $\lambda_{Dh}$, $\lambda_{DH}$.
In order to compute the DM relic density, we have solved the  
Boltzmann equation (Eq. (\ref{be1})) numerically using
the micrOMEGAs \cite{Belanger:2013oya} package
while the information of the present model is supplied to micrOMEGAs
through the LanHEP \cite{Semenov:2010qt} package. All the constraints on the model
parameters, listed in Section \ref{model}, are also taken into account
in the numerical calculations. 

\begin{figure}[]
\centering
\includegraphics[angle=0,height=7cm,width=8cm]{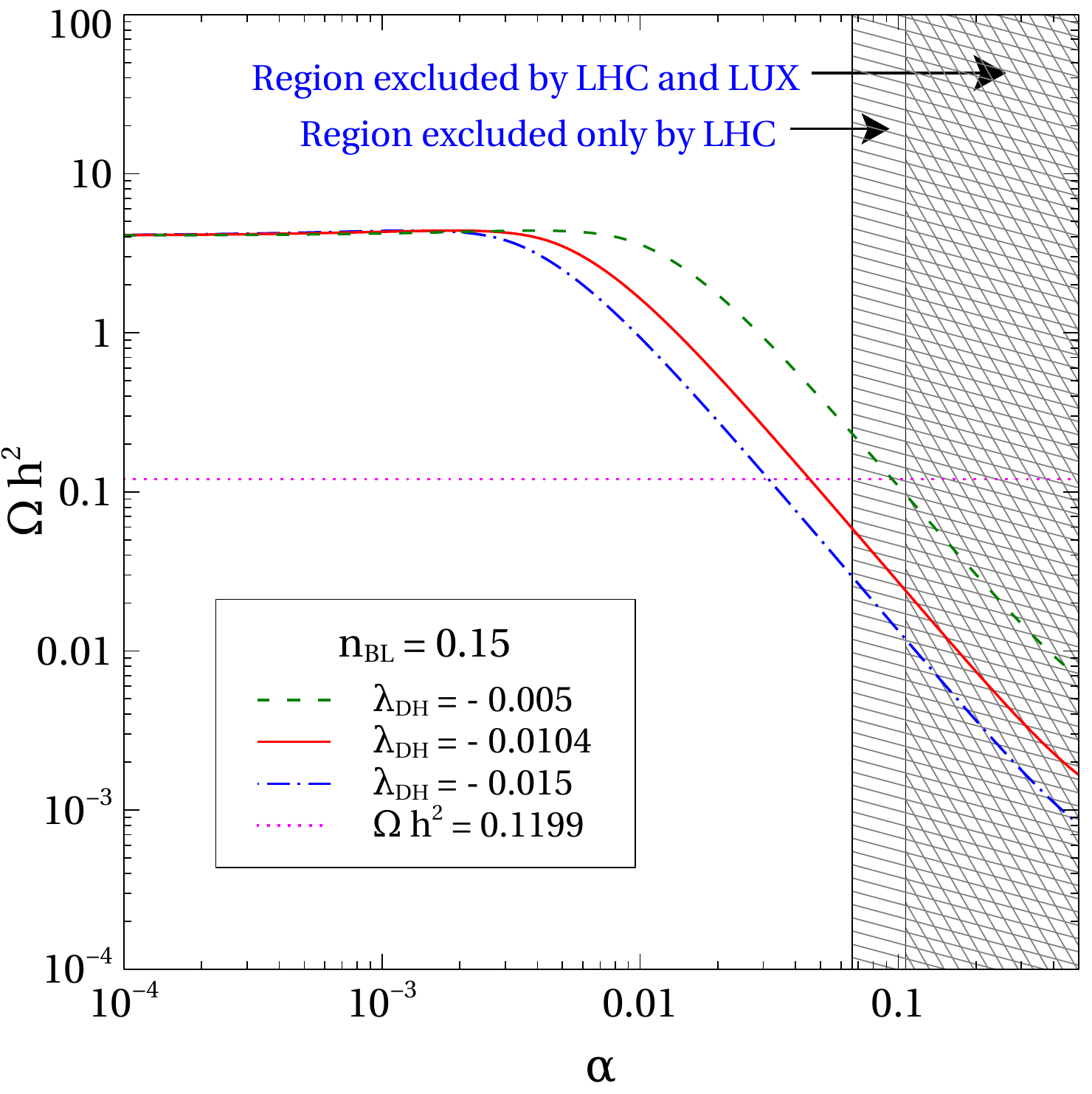}
\includegraphics[angle=0,height=7cm,width=8cm]{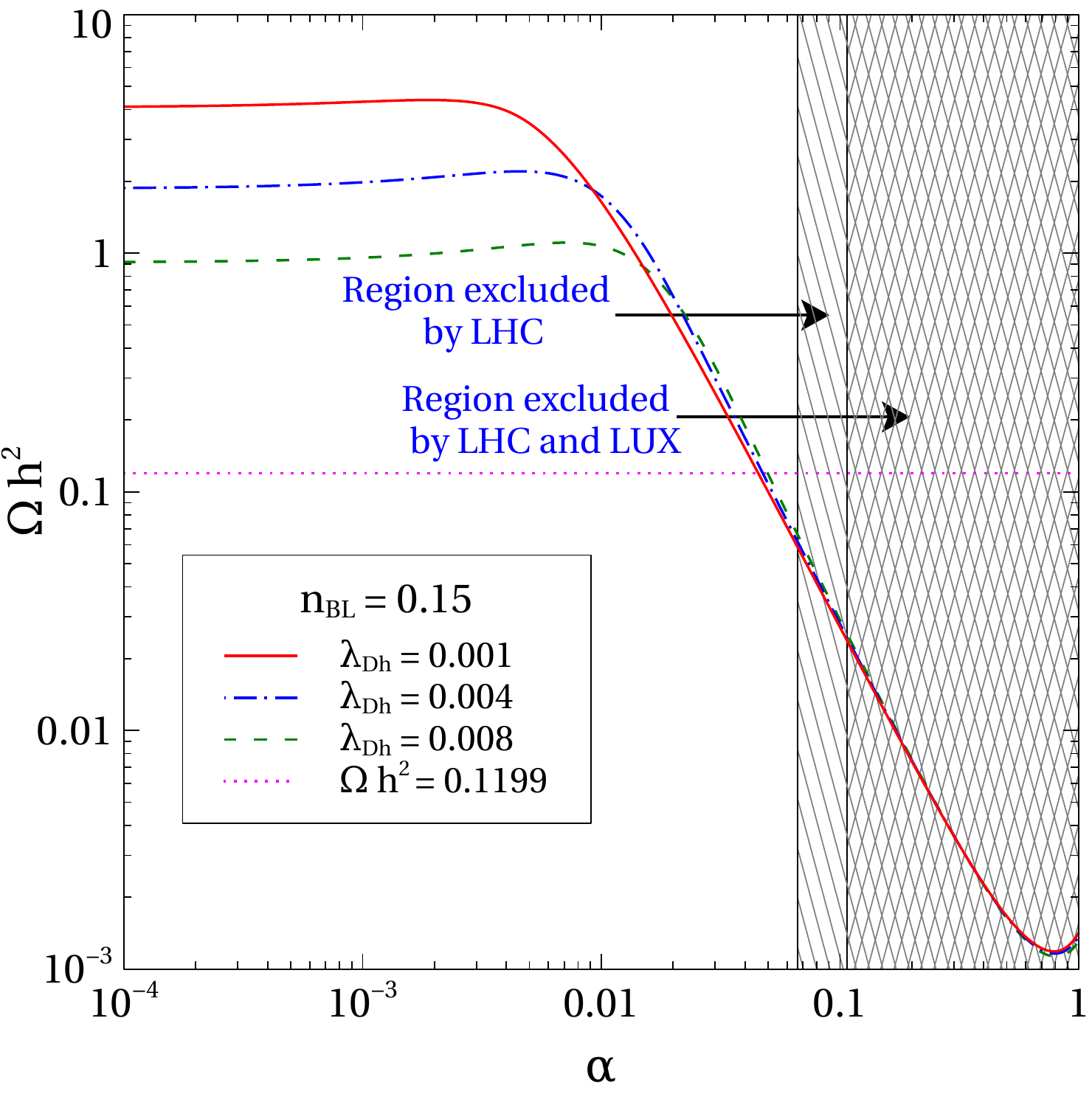}
\caption{Left (Right) panel: Variation of relic density $\Omega h^{2}$
with mixing angle $\alpha$ for $n_{BL} = 0.15$
and three different values of $\lambda_{DH}$ ($\lambda_{Dh}$) while other parameters
value have been kept fixed at $g_{BL} = 0.01$, $M_{DM} = 52.0$ GeV,
$M_{h_2} = 102.8$ GeV, $M_{Z_{BL}} = 104.1$ GeV, $\lambda_{Dh}
= 0.001$ ($\lambda_{DH} = -0.0104$).}
\label{1a}
\end{figure}

In the left panel (right panel) of Fig. \ref{1a} we plot
the variation of DM relic density $\Omega h^2$ with the scalar
mixing angle $\alpha$ for three different values of
$\lambda_{DH}=-0.005$ ($\lambda_{Dh} = 0.008$)
(green dashed line), $-0.0104$ ($0.001$) (red solid line) and $-0.015$ ($0.004$)
(blue dashed-dotted line) while the values of other parameters are kept
fixed at $g_{BL}=0.01$, $M_{DM}=52.0$ GeV, $M_{h_2}=102.8$ GeV, $M_{Z_{BL}}=104.1$ GeV,
$\lambda_{Dh}=0.001$ ($\lambda_{DH} = -0.0104$) and $n_{BL} = 0.15$.
In this plot, magenta dotted line
represents the central value of DM relic density as reported by
the Planck collaboration ($\Omega h^{2} = 0.1199$). 
From the Table \ref{tab2} we see that the $Z_{BL}$ mediated diagram is 
independent of the mixing angle $\alpha$, so its contribution does not 
depend on $\alpha$. On the other hand the two Higgs scalars 
$h_1, h_2$ mediated diagrams are dependent on the mixing angle $\alpha$. 
It is seen from Fig. \ref{1a} that the dark matter 
relic density is practically independent of the mixing angle $\alpha$
when $\alpha$ becomes too small ($\alpha < 3 \times 10^{-3}$). This can be 
explained as follows, 
in this region $\sin\alpha \sim 0$ and the $h_2$ mediated diagram does not contribute 
since the $l\,\bar{l}\,h_2$ vertex is suppressed and it is mostly the $Z_{BL}$ and  $h_1$  mediated diagrams that contribute. For very small $\alpha$, even the 
$h_1$ mediated diagram is independent of $\alpha$ since $\cos\alpha \sim 1$, 
making $\Omega h^{2}$ constant with $\alpha$. We also note from
left (right) panel of Fig. \ref{1a}
that in this region, $\Omega h^{2}$ has no 
dependence on $\lambda_{DH}$ ($\lambda_{Dh}$). This again can be 
explained using the fact that here only the $h_1$ mediated diagram 
(in addition to the $Z_{BL}$ mediated diagram which is anyway independent 
of $\alpha$, $\lambda_{DH}$ and $\lambda_{Dh}$) contributes and 
Table \ref{tab2} reveals that for small $\alpha$ we have impact of only $\lambda_{Dh}$ 
on $\Omega h^{2}$. 

On the other hand if we start increasing the mixing $\alpha$ after the value 
($\alpha > 3 \times 10^{-3}$), the scalars $h_1$ and $h_2$ both start contributing 
along with ${\rm B-L}$ gauge boson $Z_{BL}$
in the DM annihilation process, which enhances
${\langle \sigma {\rm v} \rangle}_{b \bar{b}}$. Therefore
the relic density which is approximately inverse of
${\langle \sigma {\rm v} \rangle}_{b \bar{b}}$ decreases with
increase of mixing angle $\alpha$. 
Again we notice from Table \ref{tab2} that the $\cos\alpha$ dependent term in 
the vertex $\phi_{DM} \phi_{DM}^{\dagger} h_2$ and the $\sin\alpha$ dependent 
term in the vertex $\phi_{DM} \phi_{DM}^{\dagger} h_1$ is proportional to 
$\lambda_{DH}$. This makes the relic density decrease with increasing 
$\lambda_{DH}$ 
for larger values of $\alpha$, as is evident from the left panel of Fig. \ref{1a}. 
Likewise the right panel shows the dependence of the relic density on 
$\lambda_{Dh}$ which comes from the first term of the $\phi_{DM} \phi_{DM}
^{\dagger} h_1$ vertex. This explains the decrease of the relic density 
with $\lambda_{Dh}$. 
In this figure, we have also shown the excluded region for mixing angle $\alpha$ 
from LUX and LHC experiment. The crossed region is excluded by both 
LUX and LHC experiment, whereas the forward lines is only excluded by LHC experiment.


\begin{figure}[h!]
\centering
\includegraphics[angle=0,height=7cm,width=8cm]{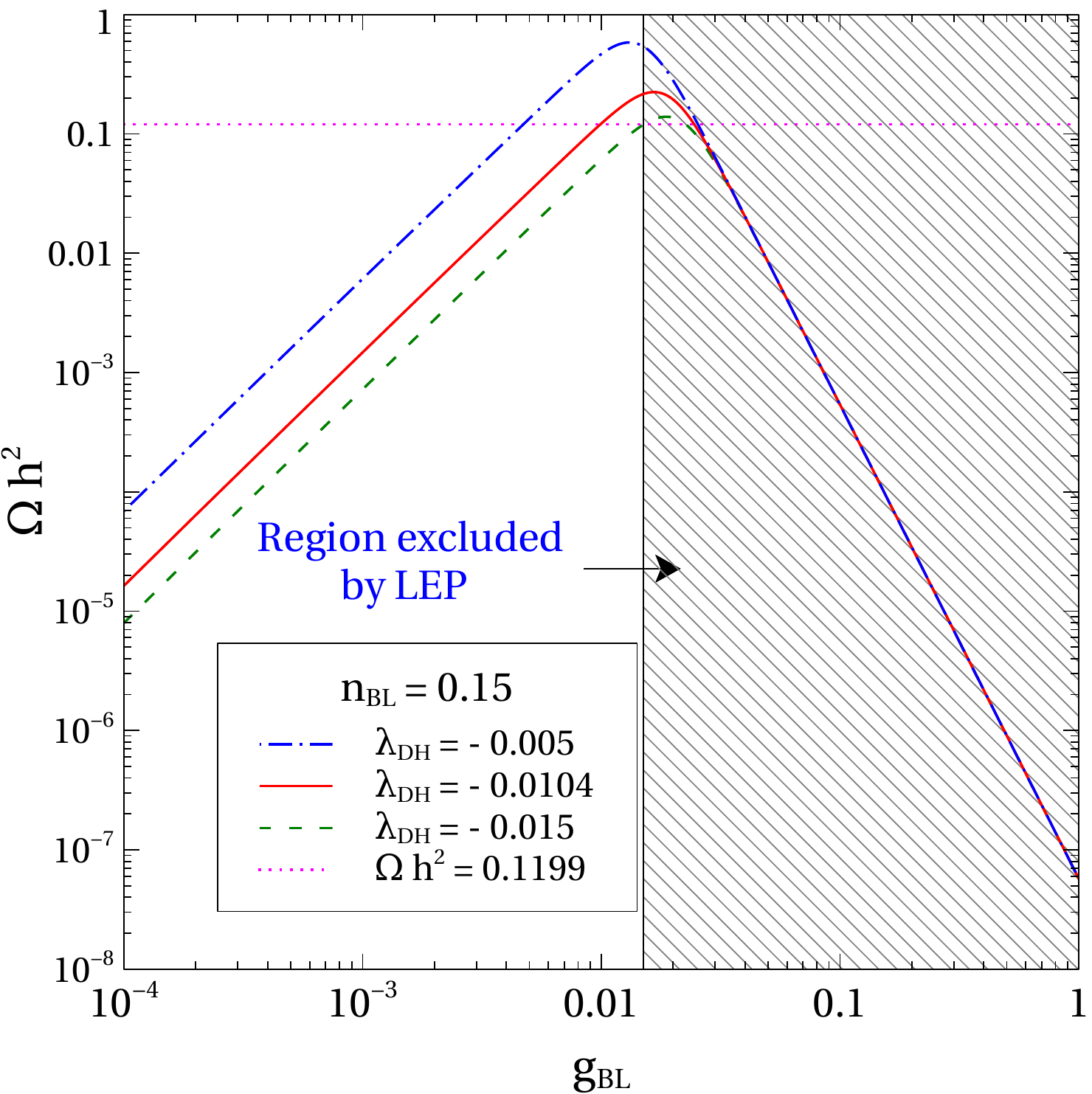}
\includegraphics[angle=0,height=7cm,width=8cm]{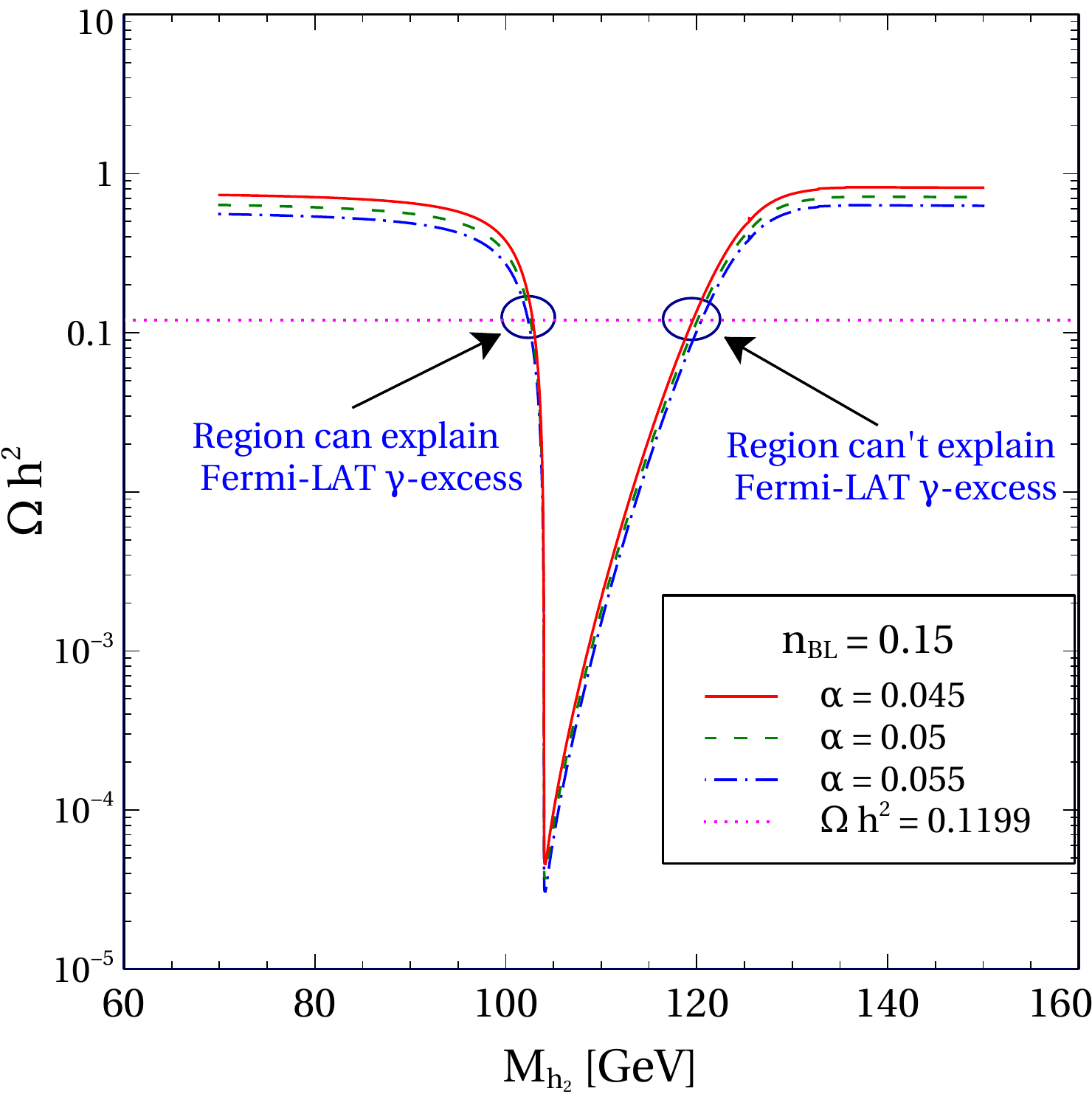}
\caption{Left (Right) panel: Variation of relic density $\Omega h^{2}$
with $g_{BL}$ ($M_{h_2}$) for $n_{BL} = 0.15$
and three different values of $\lambda_{DH}$ ($\alpha$)
while other parameters value have been kept fixed at
$M_{DM} = 52.0$ GeV, $M_{Z_{BL}} = 104.1$ GeV, $M_{h_{2}} = 
102.8$ GeV, $\lambda_{Dh} = 0.001$, $\alpha = 0.045$
($\lambda_{DH} = -0.0104$). For discussion about
the two marked regions see text below of this figure.}
\label{2a}
\end{figure}

Left panel of Fig. \ref{2a} represents the variation of $\Omega h^2$ with
$\bl$ gauge coupling $g_{BL}$ for three different chosen values of $\lambda_{DH}$.
Here green dashed-dotted line is for $\lambda_{DH} = - 0.015$, red solid line is for
$\lambda_{DH} = - 0.0104$ whereas the plot for $\lambda_{DH} = - 0.015$ is
shown by blue dashed line. Like the previous figures here also, the central value
of Planck limit on DM relic density is indicated by magenta dotted line. It is
seen from the left panel of Fig. \ref{2a} that initially the relic density increases
with $g_{BL}$ and attains a maximum value at $g_{BL}\sim 0.01$, thereafter it starts 
decreasing with $g_{BL}$. The initial rise of $\Omega h^2$, for low $g_{BL}$,
is due to s channel process of $\phi_{DM} \phi_{DM}^\dagger \rightarrow f \bar{f}$,
mediated by $h_1$ and $h_2$. In this case, the relevant couplings
($\phi_{DM}^{\dagger} \phi_{DM} h_i$, $i=1,$ 2) are
inversely proportional to $g_{BL}$ (see Table \ref{tab2}). However,
as $g_{BL}$ becomes large ($g_{BL} \ga 0.01$),  the other s channel process
mediated by ${\rm B-L}$ gauge boson starts dominating over the scalar
exchange processes. From Table \ref{tab2}, one can easily see that the coupling
$\phi_{DM}^{\dagger} \phi_{DM} Z_{BL}$ is proportional to $g_{BL}$,
which makes ${\langle \sigma {\rm v} \rangle}_{f \bar{f}}$ (via
$Z_{BL}$ exchange) proportional to fourth power of
$g_{BL}$ \footnote{$f \bar{f} Z_{BL}$ coupling is also proportional to $g_{BL}$.}.
The dominance of s channel $Z_{BL}$ exchange annihilation process
over the scalar mediated ones is indicated by the fact that in this region
(higher value of $g_{BL}$, $g_{BL} \ga 0.01$) $\Omega h^2$
(or ${\langle \sigma {\rm v} \rangle}_{f \bar{f}}$) does not depend
on the coupling $\lambda_{DH}$. We show by the hatched region the values of 
$g_{BL}$ excluded by LEP. 

\begin{figure}[h!]
\centering
\includegraphics[angle=0,height=8cm,width=10cm]{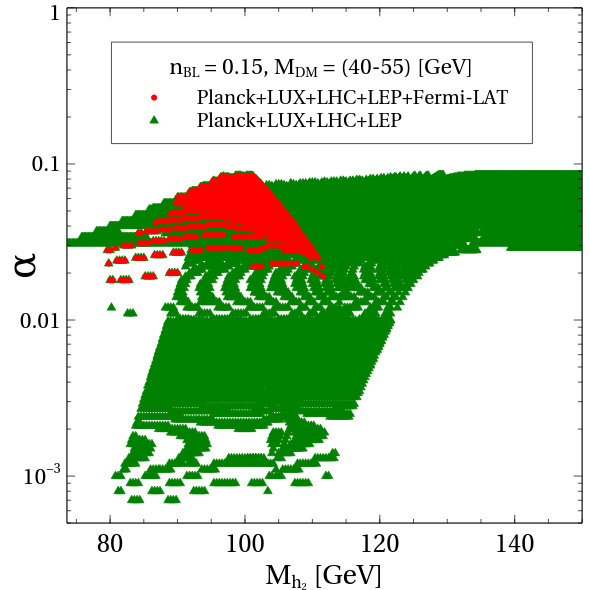}
\caption{Allowed region in $M_{h_2}$-$\alpha$ plane satisfied by various
experimental constraints considered in this work. Other relevant
parameters are kept fixed at $\lambda_{Dh}=0.001$, $\lambda_{DH}=-0.0104$,
$M_{Z_{\rm BL}}=104.1$ GeV and $g_{\rm BL}=0.01$.}
\label{7a}
\end{figure}

In the right panel of Fig. \ref{2a} we 
show the variation of $\Omega h^2$ with the mass of the non-standard Higgs
boson $h_2$ for three different values of its mixing angle with SM Higgs, 
namely $\alpha = 0.045$, 0.05, 0.055. From this plot, it is seen
that for all the chosen values of $\alpha$ the relic density satisfies
the Planck limit only near the resonance region when $M_{DM} \sim M_{h_2}/2$. 
The figure shows that in this region 
$\Omega h^2$ becomes practically independent of $\alpha$. We see that there 
are two sets of values of $M_{h_2}$ for which the model can predict the correct 
dark matter relic density. 
Of these two regions which are marked in the figure, one of them
with $M_{h_2}\sim 100$ GeV produces
$\langle{\sigma {\rm v}}\rangle_{b\bar{b}}$ in the right ballpark
value of $\sim 10^{-26}\,{\rm cm^3}/{\rm s}$, thus
can explain the Fermi-LAT gamma-ray excess \cite{Calore:2014nla}.
Whereas the other region labelled as
``Region can't explain Fermi-LAT $\gamma$ excess''
($M_{h_2}\sim 120\,\,{\rm GeV}$) produces
$\langle{\sigma {\rm v}}\rangle_{b\bar{b}} \sim 10^{-29}\,{\rm cm^3}/{\rm s}$
(see Fig. \ref{7a} also).
\begin{figure}[]
\centering
\includegraphics[angle=0,height=7cm,width=8cm]{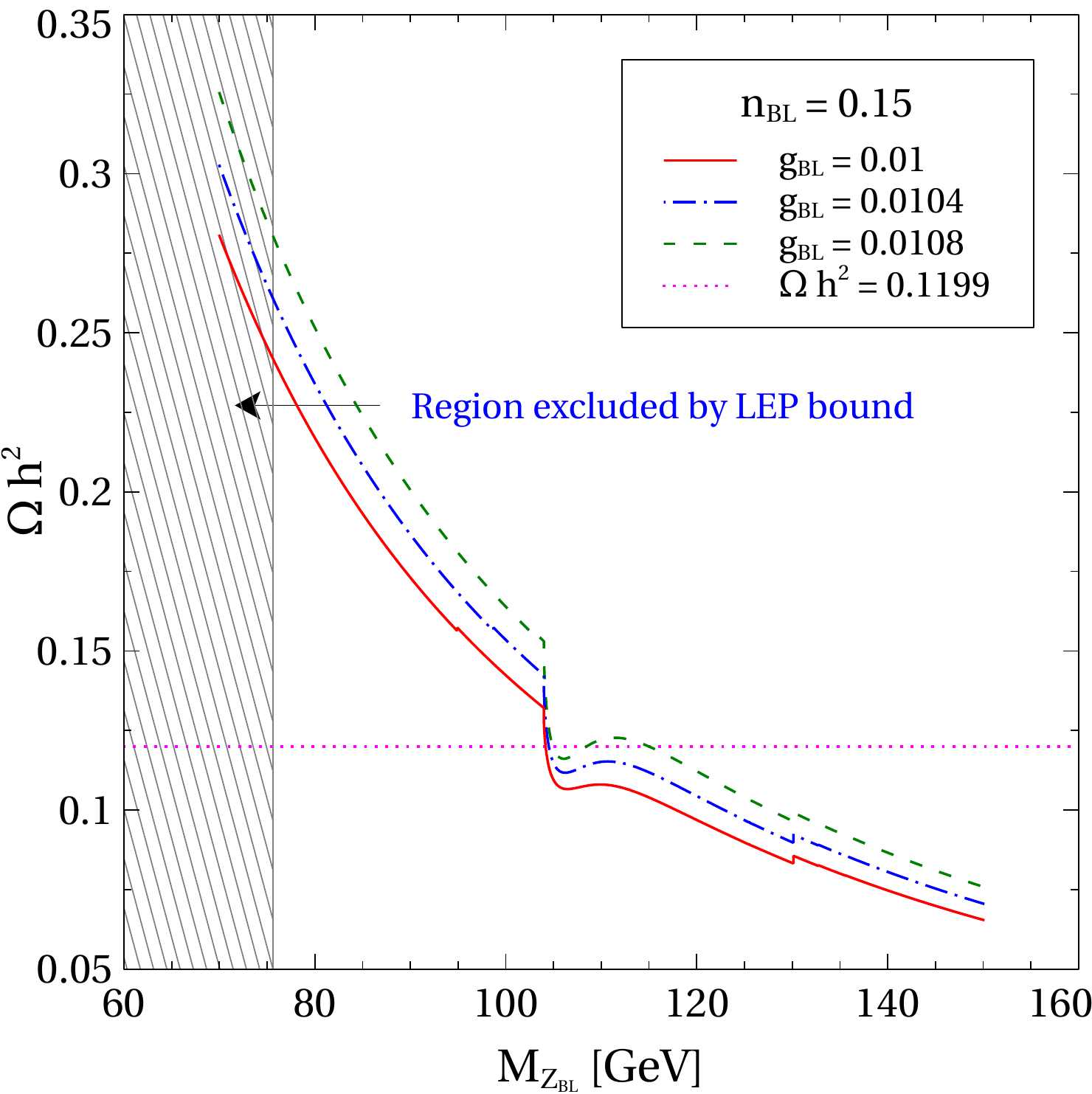}
\includegraphics[angle=0,height=7cm,width=8cm]{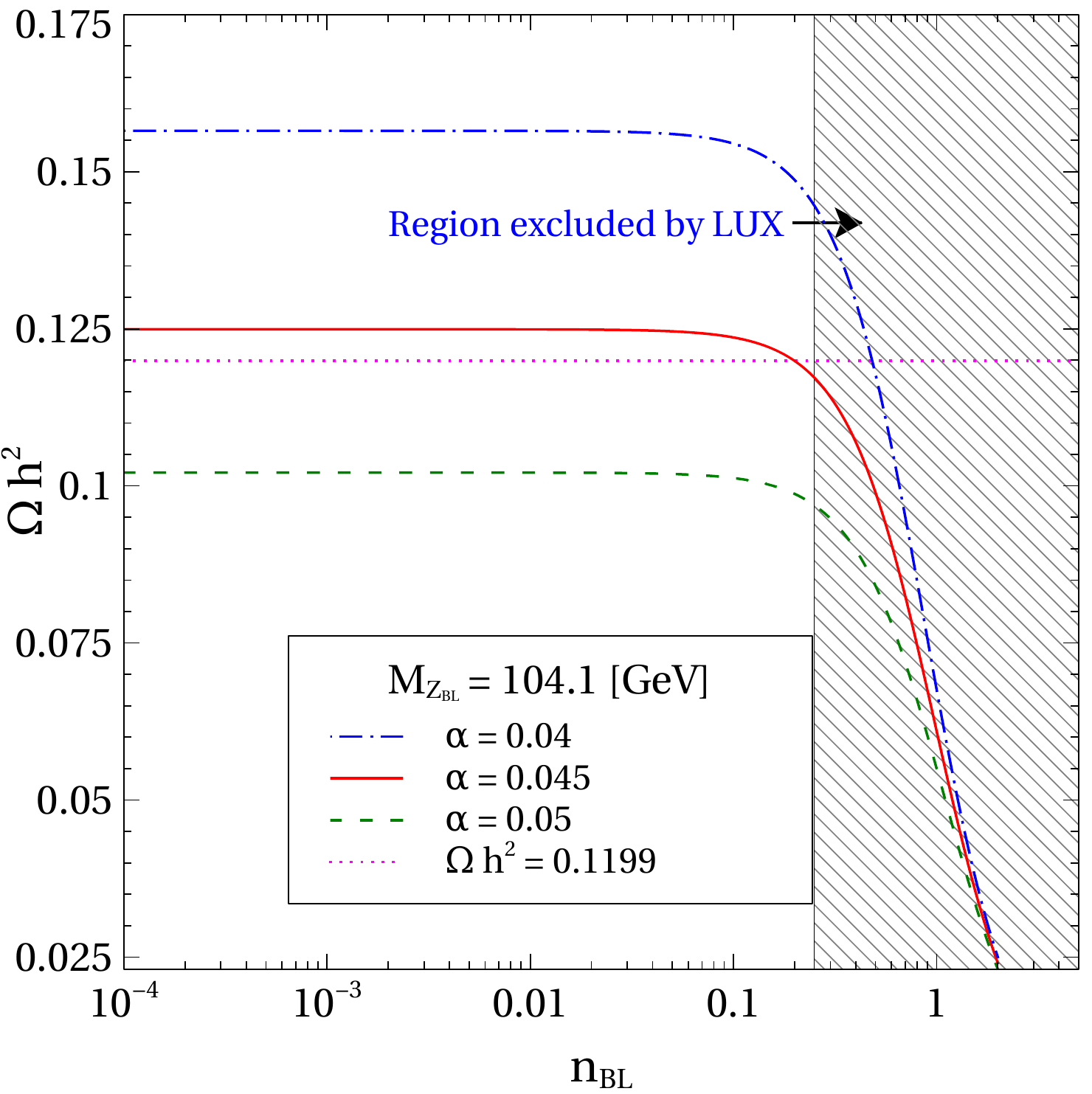}
\caption{Left panel: Variation of relic density $\Omega h^{2}$
with the mass of $Z_{BL}$ for three different values of
$g_{BL}$. Right panel: Variation of DM relic density with its ${\rm B-L}$
gauge charge $n_{BL}$ for three different values of $\alpha$.
Both the plots are drawn for $M_{DM} = 52.0$ GeV,
$M_{h_2} = 102.8$ GeV, $\lambda_{Dh} = 0.001$, $\lambda_{Dh} = - 0.0104$.}
\label{3a}
\end{figure} 

Variation of $\Omega h^2$ with the mass of ${\rm B-L}$ gauge boson
is shown in left panel of Fig. \ref{3a}. In this figure three different
plots are computed for three different values of ${\rm B-L}$ gauge coupling ($g_{BL}$).
Here, red solid line is for $g_{BL} = 0.01$ while $g_{BL} = 0.0108$ and
$0.0104$ are represented by green dashed line and blue 
dashed-dotted line, respectively.
This figure is drawn for fixed values of other parameters,
namely, $\alpha = 0.045$, $M_{DM} = 52.0$ GeV, $M_{h_2} = 102.8$ GeV,
$\lambda_{Dh} = 0.001$, $\lambda_{DH} = - 0.0104$, $n_{BL} = 0.15$. 
From this plot it is seen that for a fixed value of $M_{Z_{BL}}$, DM relic density
increases with $g_{BL}$, which is consistent with the plot in left panel
of Fig. \ref{2a} (cf. red line in the left panel of Fig. \ref{2a}
where the maxima of $\Omega h^2$ occurs for $g_{BL} \ga 0.015$). The presence of
resonance due to $Z_{BL}$ (when $\sqrt{s} \simeq M_{Z_{BL}}$) is also
seen from this figure and like the previous case for
$h_2$ here also the Planck limit is satisfied only near the resonance. However,
the resonance due to $Z_{BL}$ is not as sharp as it is due to $h_2$ because
in this region of parameter space the decay width of $Z_{BL}$
is nearly two orders of magnitude larger than that of $h_2$. Here in the left panel 
the shaded region is not allowed by the LEP bound on $Z_{BL}$. 
Right panel of Fig. \ref{3a} describes the variation of
$\Omega h^2$ with the $B-L$ gauge charge ($n_{BL}$) for
three different values of neutral scalar mixing angle
namely $\alpha = 0.04$ (blue dashed-dotted line), $0.045$ (red solid line)
and $0.05$ (green dashed line), respectively. From this figure
it is seen that as the ${\rm B-L}$ charge of the DM candidate $\phi_{DM}$
decreases, its relic density increases sharply and eventually the DM
relic density saturates after a certain value of $n_{BL} \la 0.1$.
A possible explanation of this nature of $\Omega h^2$ could be
as follows. For large value of $n_{BL}$ ($n_{BL} \sim 1$) the
maximum contribution to DM annihilation cross section
comes from ${\rm B-L}$ gauge boson mediated channel as the
cross section for this channel is directly proportional to $n^2_{BL}$.
Hence ${\langle \sigma {\rm v} \rangle}_{f \bar{f}}$ becomes
practically independent of the mixing angle $\alpha$. However
as $n_{BL}$ decreases from unity the scalar mediated s channel processes
become significant and consequently after a certain value of
$n_{BL}$ ($n_{BL}\la 0.1$) the annihilation cross section ${\langle \sigma {\rm v}
\rangle}_{f \bar{f}}$ becomes nearly insensitive to $n_{BL}$ and depends strongly
on the mixing angle $\alpha$. In the right panel,
we have given upper bound on the DM charge $n_{BL}$,
which we get from LUX limit on spin
independent direct detection cross section.  
   
\begin{figure}[h!]
\centering
\includegraphics[angle=0,height=7cm,width=8cm]{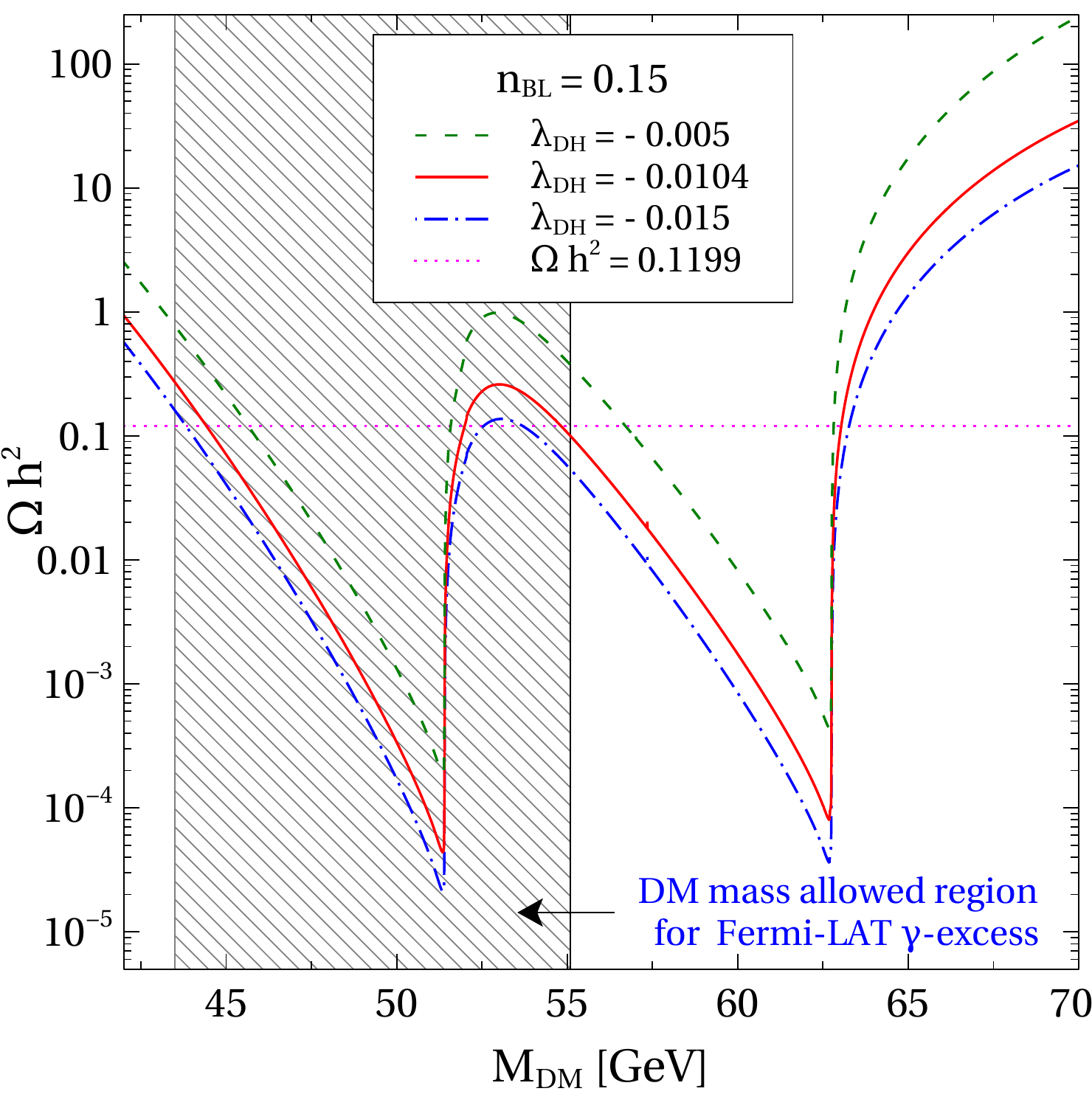}
\includegraphics[angle=0,height=7cm,width=8cm]{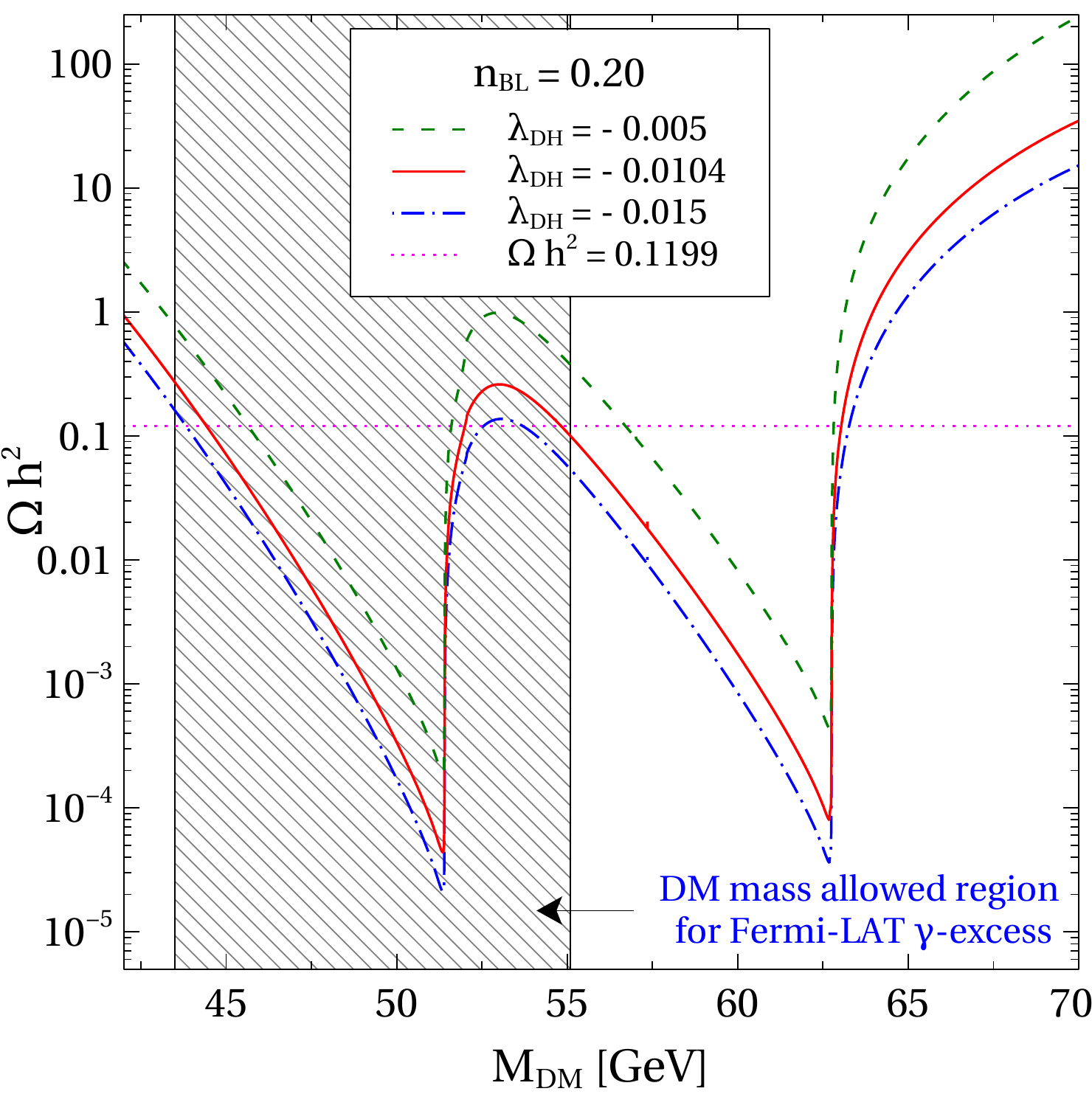}
\caption{Left (Right) panel: Variation of relic density $\Omega h^{2}$
with mass of $\phi_{DM}$ for $n_{BL} = 0.15$ ($n_{BL} = 0.20$)
and three different value of $\lambda_{DH}$ while other parameters
value have been kept fixed at $\alpha = 0.045$, $g_{BL} = 0.01$,
$M_{h_2} = 102.8$ GeV, $M_{Z_{BL}} = 104.1$ GeV, $\lambda_{Dh} = 0.001$.}
\label{4a}
\end{figure}
 
Variation of $\Omega h^2$ with dark matter mass for two different
values of $n_{BL}$ are shown in Fig. \ref{4a}. In this
figure, the left panel is for $n_{BL} = 0.15$ while the right panel is for $n_{BL}=0.2$.
In each panel the three different lines
represent the variation of $\Omega h^2$ with $n_{BL}$ for three
chosen values of $\lambda_{DH} = -0.005$, $-0.0104$ and $-0.015$ respectively.
From both panels of Fig. \ref{4a} it is seen that there are two resonance
regions where the first one is for the non-standard Higgs boson $h_2$
($M_{h_2} \sim 104$ GeV) while the second one corresponds to the SM Higgs
boson of mass 125.5 GeV. In both panels the DM relic density
satisfies the Planck limit (indicated by the magenta dotted line)
only near the resonance regions. In both the panel of Fig. \ref{4a},
we have shown allowed region of DM mass for explaining Fermi-LAT
gamma-ray excess from GC.

\begin{figure}[h!]
\centering
\includegraphics[angle=0,height=7cm,width=8cm]{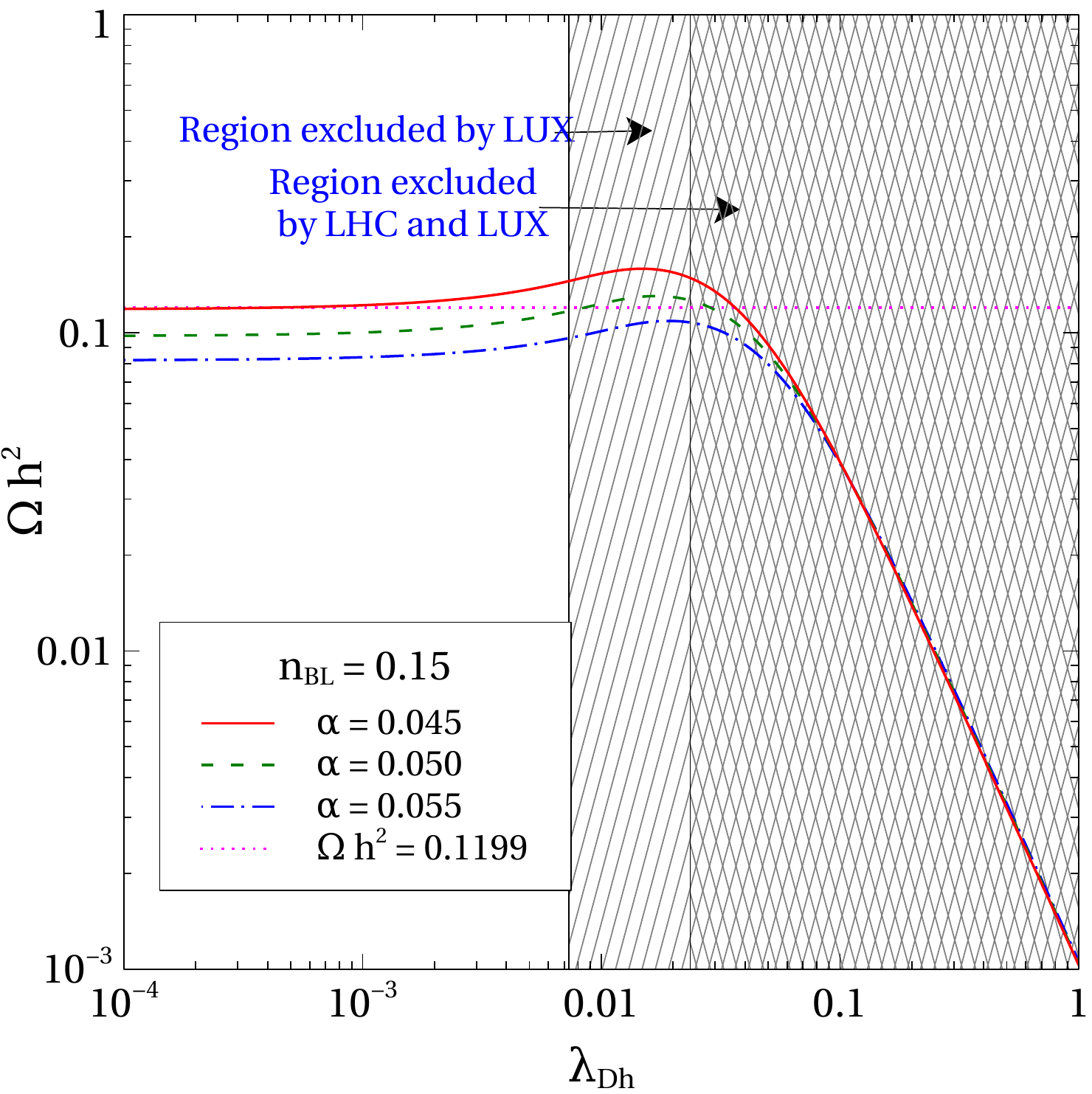}
\includegraphics[angle=0,height=7cm,width=8cm]{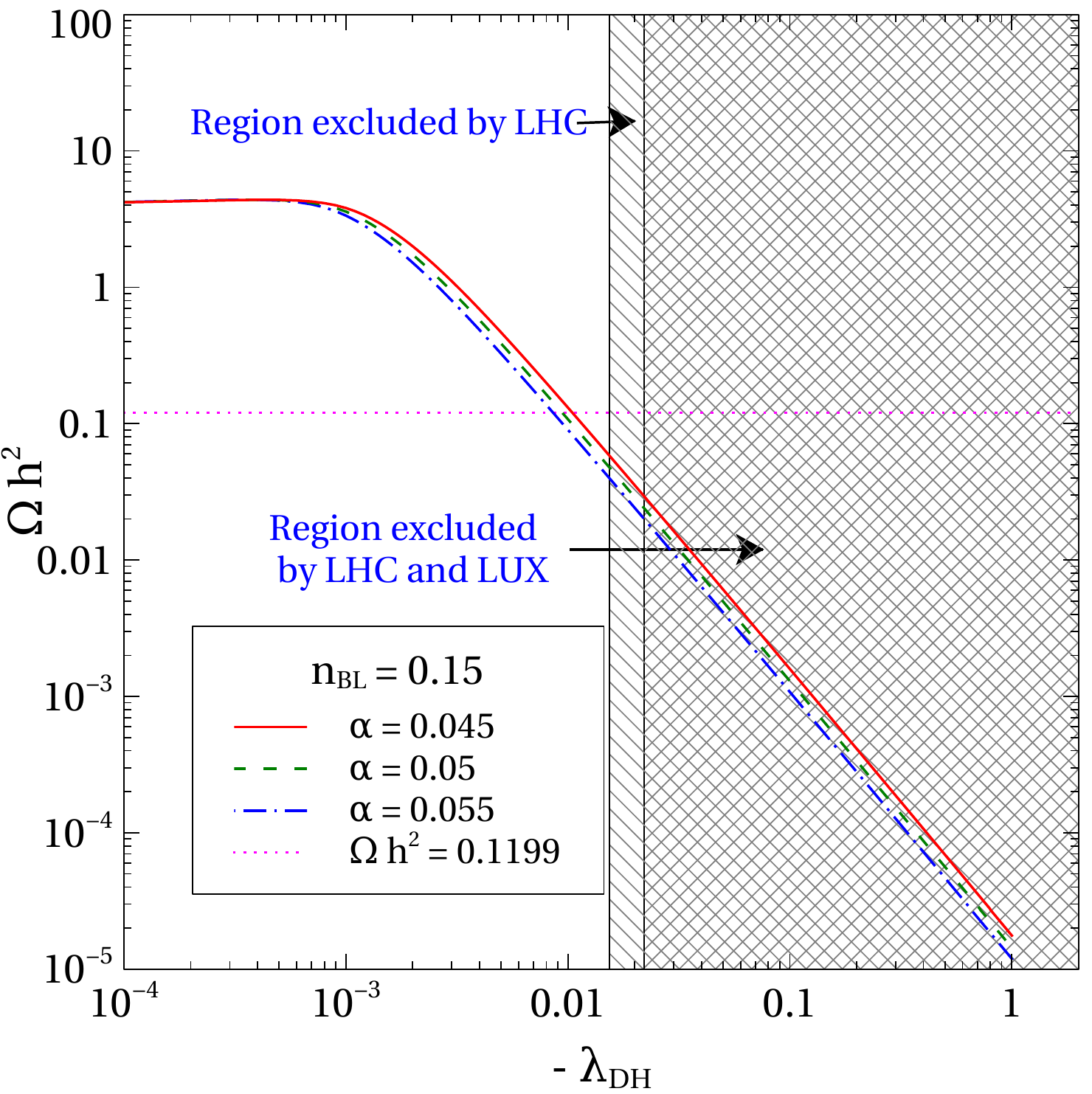}
\caption{Left (Right) panel: Variation of relic density
$\Omega h^{2}$ with $\lambda_{Dh}$ ($\lambda_{DH}$) for $n_{BL} = 0.15$
and three different values of mixing anle $\alpha$ while
other relevant parameters value have been kept fixed at $M_{DM} = 52.0$ GeV,
$M_{h_2} = 102.8$ GeV, $M_{Z_{BL}} = 104.1$ GeV,
$\lambda_{DH} = - 0.0104$ ($\lambda_{Dh} = 0.001$).}
\label{5a}
\end{figure}

We finally show the variation of $\Omega h^2$ with two
remaining model parameters $\lambda_{Dh}$ and $\lambda_{DH}$
in left and right panel of Fig. \ref{5a}, respectively. In each panel
we have shown the variation of $\Omega h^2$ for three different
values of mixing angle $\alpha$ namely $\alpha = 0.045$, $0.05$ and $0.055$.
From the left panel of Fig. \ref{5a} it is seen that for small value of
the parameter $\lambda_{Dh}$ ($\lambda_{Dh}<0.03$) relic density remains
unaffected with respect to the change in value of $\lambda_{Dh}$ as
in this region DM annihilation cross section is controlled by the coupling
$\lambda_{DH}$ which is considered to be $|\lambda_{DH}| \sim 0.01$. Also
from Table \ref{tab2} we see that
when $\lambda_{DH} \gg \lambda_{Dh}$, the couplings
$g_{h_1 \phi_{DM} \phi^{\dagger}_{DM}} \propto \sin \alpha$ and
$g_{h_2 \phi_{DM} \phi^{\dagger}_{DM}} \propto \cos \alpha$. However,
the term within the modulus in Eq. (\ref{sigmaff}) is proportional to
$\sin^2 \alpha$. Therefore, inspite of being small in value, the variation
of $\alpha$ produces a significant change in $\sigma$ and hence in
relic density. Similarly, using Eq. (\ref{sigmaff}) and Table \ref{tab2}
one can easily see that for higher value of $\lambda_{Dh}$ (when
$\lambda_{DH} \ll \lambda_{Dh}$), the scalar mediated term in $\sigma$
(term within modulus in Eq. (\ref{sigmaff})) mainly depends on
$\cos \alpha$ and $\lambda_{Dh}$. Consequently, for the higher
value of $\lambda_{Dh}$, there is no observable change in relic density
with respect to $\alpha$ and it decreases with the increase of $\lambda_{Dh}$.
In right panel of Fig. \ref{5a} we have shown the variation of $\Omega h^2$
with $\lambda_{DH}$. It is seen from this figure that, 
the behaviour of DM relic density with respect to the coupling $\lambda_{DH}$
is same as it is with $\lambda_{Dh}$ i.e. initially for small value of
$\lambda_{DH}$ relic density remains unchanged and therefore after a
certain value of $\lambda_{DH}$ (when $\lambda_{DH} > \lambda_{Dh}$,
$\lambda_{Dh} \sim 10^{-3}$) relic density falls gradually with the increase
of $\lambda_{DH}$. However, by comparing both the plots in Fig. \ref{5a}
one finds that with respect to $\alpha$ the behaviour of
$\Omega h^2$ Vs $\lambda_{DH}$ curve is exactly opposite to
the curve $\Omega h^2$ Vs $\lambda_{Dh}$ (shown in the left panel)
which can be easily understood from Table \ref{tab2} and Eq. (\ref{sigmaff}).
In both the panel we have shown allowed regions for the coupling constant
$\lambda_{Dh}$ and $\lambda_{DH}$ respectively. The crossed regions are
excluded by both LHC and LUX, whereas for left panel the backward line
region is excluded by LUX and for right panel the forward line region
is excluded by LHC.   

\begin{figure}[h!]
\centering
\includegraphics[angle=0,height=7cm,width=8cm]{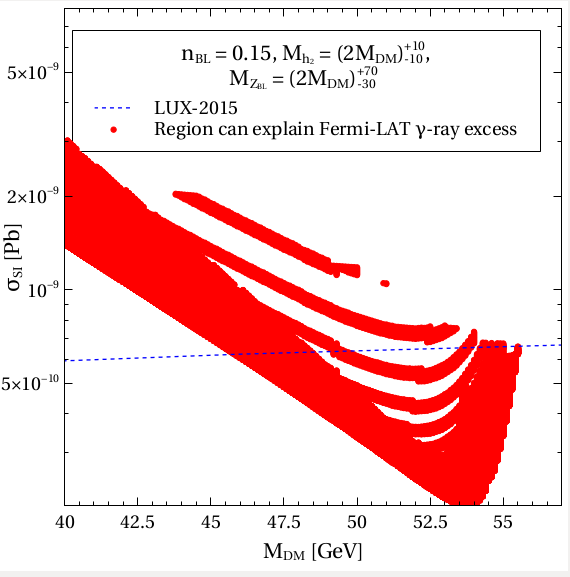}
\includegraphics[angle=0,height=7cm,width=8cm]{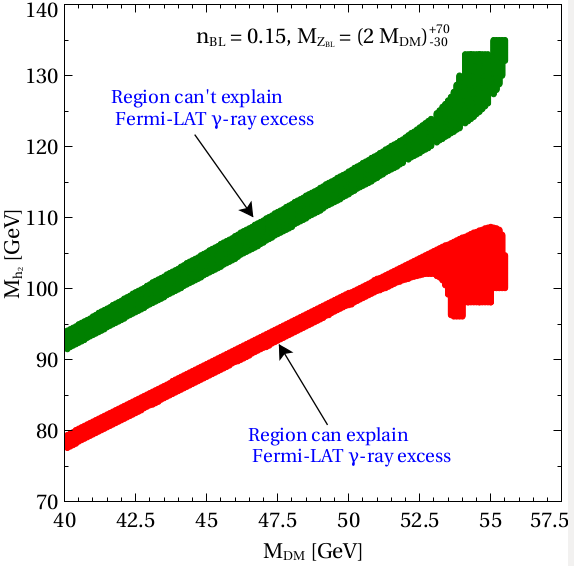}
\caption{Left panel: Spin independent cross section $\sigma_{SI}$
between and dark matter particle ($\phi_{DM}$) and the detector
nucleon for $n_{BL} = 0.15$. Blue dashed lines in this panel 
represent upper limit on $\sigma_{SI}$ reported by
LUX collaboration.
Right panel: Allowed regions in $M_{DM}-M_{h_2}$ plane
which satisfy the observed relic density, Fermi-LAT gamma-ray excess
($\langle \sigma v \rangle_{b \bar{b}} \sim 10^{-26}$
cm$^3/$s for red coloured region only)
and LHC constraints listed in Section \ref{model}.}
\label{6a}
\end{figure}   

In the left panel of Fig. \ref{6a}, we show
how the average value of spin independent scattering cross section
$\frac{1}{2}(\sigma_{\phi_{DM}} + \sigma_{\phi^{\dagger}_{DM}})$ of
$\phi_{DM}$ and $\phi_{DM}^{\dagger}$ with the detector nuclei varies
as a function of dark matter mass for $n_{BL} = 0.15$.
While computing this 
plot, we have varied the mass of ${\rm B-L}$ gauge
boson in range of $2\,{M_{DM}}^{+ 70}_{-30}$ GeV
for a particular value of DM mass ($M_{DM}$) since the relic density
is satisfied only near the respective resonance regions of $Z_{BL}$ and $h_2$
where $M_{h_2},\,M_{Z_{BL}} \sim 2 M_{DM}$ (see Figs. \ref{2a}, \ref{3a}).
The other relevant parameters are kept fixed at $\alpha = 0.045$,
$g_{BL} = 0.01$, $\lambda_{DH} = -0.0104$, $\lambda_{Dh} = 0.001$.
The experimental upper limits on the DM
spin independent scattering cross section with the detector
nuclei is also shown by blue dashed line. Here all the points
within the red and green patches satisfy all the necessary constraints
namely Planck limit on relic density, LHC bounds on invisible decay
width and signal strength of SM-like Higgs boson ($h_1$), lower
limit on $\frac{M_{Z_{BL}}}{g_{BL}}$ from LEP and also
the vacuum stability conditions. From this plot it is
seen that although the dark matter mass between 40 GeV to 55 GeV
satisfies all the constrains mentioned above, the lower
mass region between 40 GeV to 45 GeV has already been excluded
by the upper limit on spin independent scattering cross section
reported by the LUX collaboration. Therefore in this model with
the considered ranges of model parameters, dark matter
mass of 45 GeV to 55 GeV is still allowed by all possible
experimental as well as theoretical constraints. This allowed
region can be tested in near future by the upcoming ``ton-scale''
direct detection experiments like XENON 1T.


As we have seen earlier in Fig. \ref{2a} (right panel), that for two values of
$M_{h_2}$ Planck's relic density central value is satisfied. If we consider
the higher value of $M_{h_2}$ ($M_{h_2}\sim120$ GeV) then the annihilation
cross section for the channel $\phi_{DM} \phi^{\dagger}_{DM} \rightarrow b \bar{b}$
comes in around $\langle \sigma v \rangle_{b \bar{b}} \sim 10^{-29}$
cm$^{3}\,{\rm s}^{-1}$, which cannot
explain Fermi-LAT gamma excess \cite{Calore:2014nla}.
On the other hand the lower value of $M_{h_2}$ ($M_{h_2}\sim100$ GeV)
produces $\langle \sigma v \rangle_{b \bar{b}}$ in the right ballpark
value of $10^{-26}$ cm$^3$ s$^{-1}$ which is required to explain the Fermi-LAT
gamma excess. To find the allowed region which can satisfy all
the constraints as mentioned in Section \ref{model} we have varied $M_{h_2}$
and $M_{Z_{BL}}$ in the ranges $ 2\,{M_{DM}} ^{+25}_{-10}$ GeV and
$ 2\,{M_{DM}} ^{+70}_{-30}$ GeV respectively. The allowed region in $M_{DM}-M_{h_2}$ 
plane is shown in the right panel of Fig. \ref{6a}. In this
plot red coloured region around $\sim 2 \times M_{DM}$ corresponds
to the lower value of $M_{h_2}$ which can explain the Fermi-LAT
$\gamma$-ray excess while the higher allowed value of $M_{h_2}$
is indicated by green coloured patch which is unable to explain
the GC $\gamma$-ray excess.
As we have discussed above, here also the region
corresponds to dark matter mass of 40 GeV to 45 GeV is ruled
out by the results of LUX direct detection experiment. The region
beyond the dark matter mass of 45 GeV satisfies all the constraints
listed in Section \ref{model}. 

\begin{figure}[h!]
\centering
\includegraphics[angle=0,height=7cm,width=8cm]{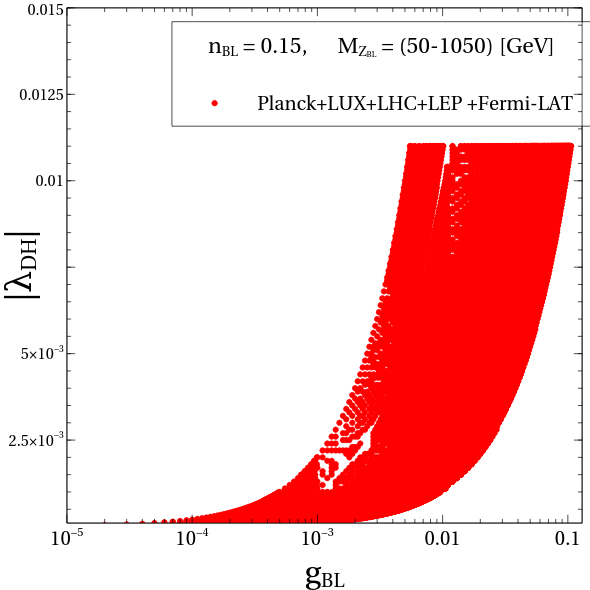}
\includegraphics[angle=0,height=7cm,width=8cm]{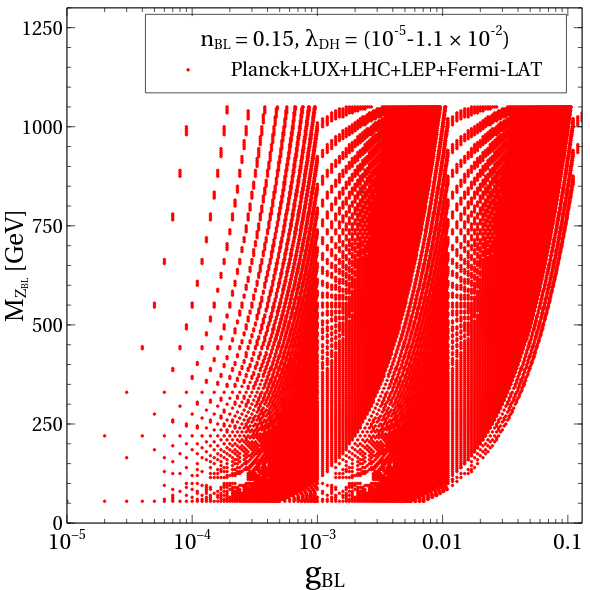}
\caption{Left panel (Right panel): Allowed region in $g_{\rm BL}$-$\lambda_{DH}$
($g_{\rm BL}$-$M_{Z_{\rm BL}}$) plane satisfied by all the experimental
constraints considered in this work. Other relevant
parameters are kept fixed at $\lambda_{Dh}=0.001$, $\alpha=0.045$,
$M_{DM}=52$ GeV and $M_{h_2}=102.8$ GeV.}
\label{8a}
\end{figure} 
In Fig. \ref{7a} we show the allowed region in $M_{h_2}$-$\alpha$ plane
for $40\,{\rm GeV}\leq M_{DM}\leq 55$ GeV, $M_{Z_{\rm BL}}=104.1$ GeV,
$n_{\rm BL} = 0.15$, $\lambda_{Dh}=0.001$, $\lambda_{DH}=-0.0104$
and $g_{\rm BL}=0.01$. Here, green coloured region satisfies 
all the constraints except Fermi-LAT bound on dark matter annihilation
cross section into $b\bar{b}$ final state while the values of $\alpha$
and $M_{h_2}$ lying within the red coloured patch are allowed by all
the experimental constraints listed in Section \ref{model}.
The region in $\lambda_{DH}-g_{\rm BL}$ plane which satisfies
simultaneously the results of Planck, LUX, LHC, LEP and Fermi-LAT experiments
is shown by a red coloured patch in the left panel of Fig. \ref{8a}. While
computing this plot we have varied the mass of the extra neutral gauge
boson $Z_{\rm BL}$ in the range of 50 GeV to 1050 GeV and the values of
other relevant parameters are kept fixed at $M_{DM} = 52$ GeV,
$M_{h_2} = 102.8\,\,{\rm GeV}$, $\alpha=0.045$, $\lambda_{Dh}=0.001$
and $n_{\rm BL} = 0.15$. From this figure it is evident that
$g_{\rm BL}\la 0.1$ and $|\lambda_{DH}|\la 0.011$ are allowed
for $50\,{\rm GeV}\leq M_{Z_{\rm BL}} \leq 1050$ GeV.
On the other hand from the right panel of Fig. \ref{8a}
one can see that all the considered
range of $M_{Z_{\rm BL}}$
($50\,{\rm GeV}\leq M_{Z_{\rm BL}} \leq 1050$ GeV), 
except the extreme right region with $g_{\rm BL}$
lies between 0.01 to 0.1 (LEP excluded region),
is allowed with respect
to the variation of U(1)$_{\rm B-L}$
gauge coupling constant $g_{\rm BL}$. 




\section{Gamma-ray flux}
\label{gammaflux}
In this present model the pair annihilation of $\phi_{DM} \phi^{\dagger}_{DM}$
produces $b$ and $\bar{b}$ at the final state\footnote{One can extrapolate
this work and can explain the Fermi-LAT gamma-ray excess by studying different
channels such as $\tau^{+} \tau^{-}$, $W^{+} W^{-}$, $q \bar{q}$ and $h_1 h_1$.}.
Therefore, these $b$ quarks
undergo hadronisation processes and produce $\gamma$-rays. The differential
gamma-ray flux from the pair annihilation of $\phi_{DM}$ and $\phi^{\dagger}_{DM}$
at the Galactic Centre region is given by      
\begin{eqnarray}
\frac{d\Phi_{\gamma}}{d \Omega d E} &=& \dfrac{1}{2} \dfrac{r_{\odot}}{8\pi}
\left(\frac{\rho_{\odot}}{M_{DM}}\right)^2 ~\bar{J}~ 
{\langle{\sigma {\rm v}} \rangle}_{b\bar{b}}
\frac{d N^{\rm b}_{\gamma}}{dE}\,\,,
\label{gamma-flux}
\end{eqnarray} 
where $r_{\odot}=8.5$ kpc is the distance of solar system from the centre
of our Milky way galaxy and dark matter density near the solar neighbourhood
is denoted by $\rho_{\odot}$ which is taken to be 0.4 GeV/cm$^3$.
Similar to Eqs. (\ref{be}, \ref{be1}), here also the half factor appearing in
the expression of the differential gamma-ray flux is due the non-self-conjugate
nature of $\phi_{DM}$. Moreover, $\frac{d N^{\rm b}_{\gamma}}{dE}$
is the spectrum of produced gamma-rays
from the hadronisation processes of $b$ quarks and we have adopted the numerical
values of $\frac{d N^{\rm b}_{\gamma}}{dE}$ for different values of photon
energy from ref. \cite{Cirelli:2010xx}. Annihilation cross section for the channel
$\phi_{DM} \phi^{\dagger}_{DM} \rightarrow b \bar{b}$ which acts
as the seed mechanism for the Galactic Centre gamma-excess, is denoted
by ${\langle{\sigma {\rm v}} \rangle}_{b\bar{b}}$.
Further, $\bar{J}$ is the averaged of
``astrophysical J factor" over a solid angle $\Delta\,\Omega$. The value
of solid angle $\Delta\,\Omega$ around the Galactic Centre depends on
the choice of a particular region of interest (ROI). In the present work
we have adopted the same ROI as considered by Calore {\it et. al.} \cite{Calore:2014nla}
which is $|l| < 20^0$ and $2^0 < |{\rm b}| < 20^0$ with $l$ and ${\rm b}$
are the galactic longitude and latitude respectively. Therefore, the expression
of $\bar{J}$ is given by
\begin{eqnarray}
\bar {J} &=& \frac{4}{\Delta \Omega} \int \int d{\rm b}~dl
\cos {\rm b}~J( {\rm b}, l) \,\,,
\label{jbar}
\end{eqnarray}
with
\begin{eqnarray}
J(l, {\rm b}) &=& \int_{\rm l.o.s} \frac{d\mathfrak{s}}{r_\odot} 
\left(\frac{\rho(r)}{\rho_{\odot}}\right)^2 \,\, ,
\label{j}
\end{eqnarray}
and
\begin{eqnarray}
\Delta\Omega = 4 \int dl \int d{\rm b} \cos {\rm b} \,\, ,
\label{solidangle} 
\end{eqnarray}
\begin{eqnarray}
r &=& \left(r_{\odot}^2 + {\mathfrak{s}}^2 - 2\,r_{\odot}\,{\mathfrak{s}}
\cos {\rm b} \cos l\right)^{{1}/{2}}\,\,\, ,
\label{radius}
\end{eqnarray}
where the integration of Eq. (\ref{j}) is performed along the line of sight
(l.o.s) distance $\mathfrak{s}$ which can be defined using Eq. (\ref{radius}).
In the definition of ``astrophysical J factor'' (Eq. (\ref{j})), $\rho(r)$
represents the variation of dark matter density with respect to the distance
$r$ from the Galactic Centre, which is also known as the density profile of
dark matter. As the actual form of the density profile is still unknown
to us there are many approximate dark matter density profiles
available in the literature such as NFW profile \cite{Navarro:1996gj},
Einasto profile \cite{einasto}, Isothermal profile \cite{Bergstrom:1997fj},
Moore profile \cite{Moore:1999gc}.
Therefore, as in ref. \cite{Calore:2014nla}, in this work also, we
have used NFW halo profile with $\gamma=1.26$, $r_s=20$ kpc. Using Eqs.
(\ref{jbar}-\ref{radius}) and a NFW dark matter halo profile we have found
the value of $\bar{J} = 57.47$ for the above mentioned ROI
($|l| < 20^0$ and $2^0 < |{\rm b}| < 20^0$). However, due to
our poor knowledge about the halo profile parameters ($\rho_{\odot}$, $\gamma$, $r_s$)
the value of $\bar{J}$ may vary from its canonical value $\bar{J} = 57.47$ obtained
for $\gamma=1.26$, $r_s=20$, $\rho_\odot = 0.4$ GeV/cm$^3$.
Now in order to include such uncertainties into the value
of $\bar{J}$, which exist within the values of DM density
profile parameters, we have redefined $\bar{J}$ in the following way
\begin{eqnarray}
\bar{J} = \mathcal{A}\,\bar{J}_{canonical} \,,
\label{errorj}
\end{eqnarray} 
where $\bar{J}_{canonical} = 57.47$, i.e. the value of $\bar{J}$ for
$\gamma=1.26$, $r_s=20$, $\rho_\odot=0.4$ GeV/cm$^3$ and the quantity
$\mathcal{A}$ can vary in the range 0.19 to 5.3 \cite{Calore:2014nla}. Therefore,
the values of $\bar{J}$ and $J_{canonical}$ coincide when $\mathcal{A}=1$.

\begin{figure}[h!]
\centering
\includegraphics[height=14cm,width=10cm,angle=-90]{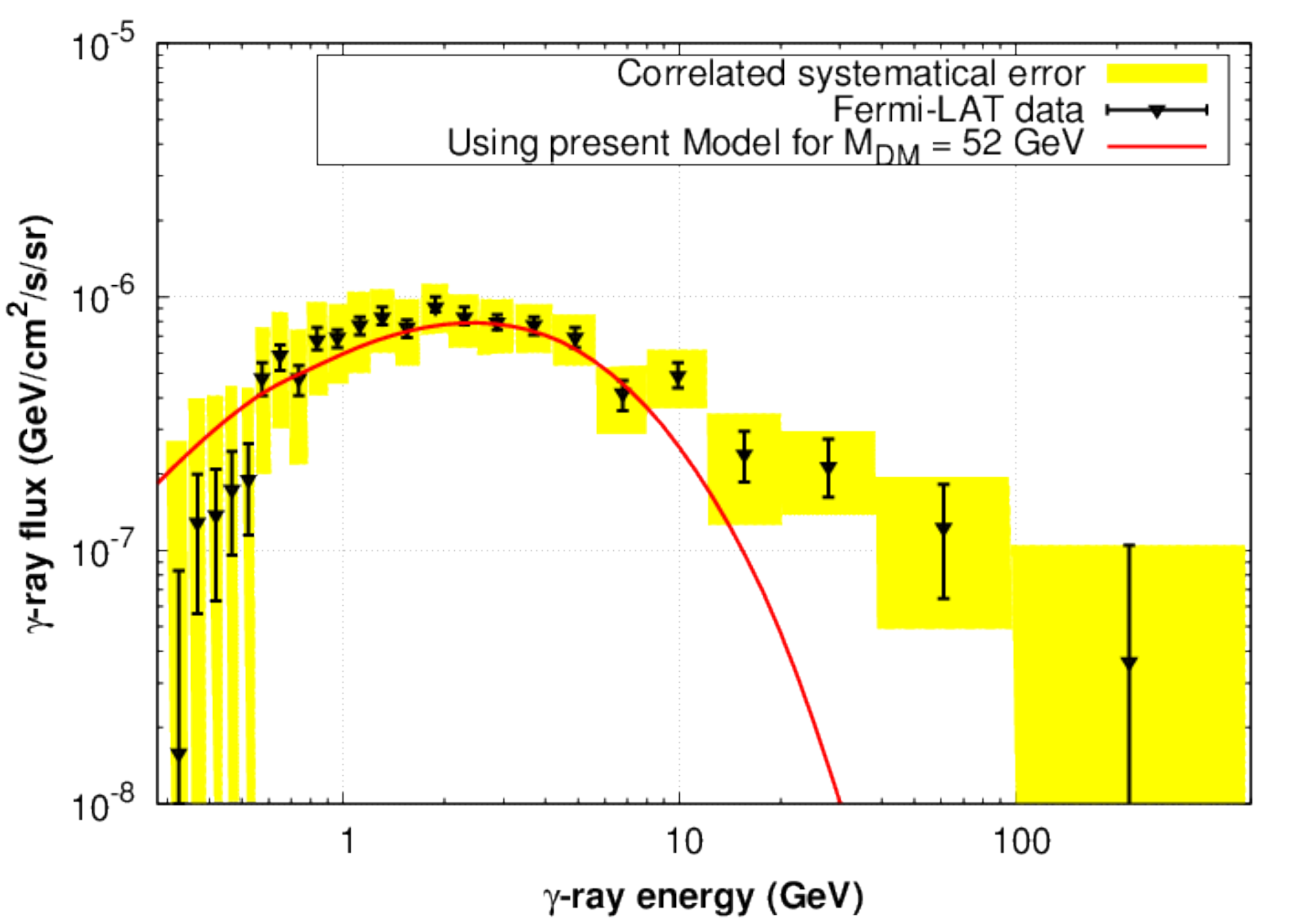}
\caption{Gamma-ray flux produced from dark matter annihilation
at the Galactic Centre.}
\label{gammaplot}
\end{figure}

Using Eqs. (\ref{gamma-flux}-\ref{errorj}), we have computed the $\gamma$-ray
flux due to the pair annihilation of $\phi_{DM} \phi^\dagger_{DM}$
into $b\bar{b}$ final state and it is plotted in Fig. \ref{gammaplot}. In this plot,
Fermi-LAT observed gamma-ray flux from the direction of Galactic Centre is
denoted by black triangle shaped points with the black vertical lines
represent the uncorrelated statistical errors while the correlated systematics
are described by yellow coloured boxes. The red solid line
denotes the gamma-ray flux which is computed for an annihilating non-self-conjugate
dark matter particle of mass $M_{DM}=52$ GeV using the present model. We have found
that the gamma-flux obtained from the present model agrees well with the
flux observed by Fermi-LAT if the product of
$\mathcal{A} \,{{\langle{\sigma {\rm v}} \rangle}_{b\bar{b}}}=4.7 \times 10^{-26}$
cm$^3/$s. Therefore, if we use the canonical values of the halo profile parameters
(when $\mathcal{A}=1$ and $\bar{J} = 57.47$) then in order to reproduced Fermi-LAT
observed gamma-ray flux from the pair annihilation of a non-self-conjugate dark matter
of mass 52 GeV its annihilation cross section for the $b\bar{b}$ channel must
be $4.7 \times 10^{-26}$ cm$^3/$s. For the other values of $\mathcal{A}$ which are not equal
to unity, the quantity ${{\langle{\sigma {\rm v}} \rangle}_{b\bar{b}}}$ will be scaled
accordingly.  

Three allowed values of ${{\langle{\sigma {\rm v}} \rangle}_{b\bar{b}}}$
that we have obtained from the present model for $M_{DM} = 52$ GeV, which
are also satisfying all the constrains listed in Section \ref{model}, are
given in Table \ref{tab3}.

\begin{center}
\begin{table}[h!]
\begin{center}
\begin{tabular}{||c|c|c|c|c|c|c||}
\hline
\hline
\begin{tabular}{c}
\small
	$M_{DM}$ [GeV]\\
	\hline
	\\
    52.0
    \\

\end{tabular}
&
\begin{tabular}{c}
\small
	$n_{BL}$\\
	\hline
	\\
	$0.15$\\

\end{tabular}
&
\begin{tabular}{c}
    \\
	$M_{h_{2}} [{\rm GeV}]$\\
	\hline
	$103.3$\\
	\hline	
	$102.8$\\
	\hline
	$101.4$\\

\end{tabular}
&
\begin{tabular}{c}
    \\
	$M_{Z_{BL}} [{\rm GeV}]$\\
	\hline
	$77.1$\\
	\hline
	$104.2$\\
	\hline
	$168.6$\\

\end{tabular}
&
\begin{tabular}{c}
    \\
	$\Omega h^{2}$\\
	\hline
	$0.1208$\\
	\hline
	$0.1191$\\
	\hline
	$0.1199$  \\

\end{tabular}
&
\begin{tabular}{c}
    \\
	$<\sigma {\rm v}>_{b \bar{b}} [{\rm cm}^{3}{\rm s}^{-1}]$\\
	\hline
	$7.005 \times 10^{-26}$\\
	\hline
	$4.545 \times 10^{-26}$\\
	\hline
	$2.853 \times 10^{-26}  $\\

\end{tabular}
&
\begin{tabular}{c}
    \\
	$\mathcal{A}$\\
	\hline
	$0.67$\\
	\hline
	$1.03$\\
	\hline
	$1.65$\\ 
    
\end{tabular}\\

\hline
\hline
\end{tabular}
\caption{Allowed values of ${{\langle{\sigma {\rm v}} \rangle}_{b\bar{b}}}$
and $\mathcal{A}$ for three randomly chosen benchmark points $M_{h_2}$
and $M_{Z_{\rm BL}}$. The values of other relevant parameters are
$g_{\rm BL}=0.01$, $\alpha=0.045$, $\lambda_{DH}=-0.0104$ and
$\lambda_{Dh}=0.001$.}
\label{tab3}
\end{center}
\end{table}
\end{center}

\section{Summary and Conclusion}
\label{summary}

Existence of neutrino masses and dark matter in the Universe are two of 
the main observational evidences for physics beyond the Standard Model. 
If the Galactic Centre gamma ray excess reported by the Fermi-LAT 
data is indeed due to 
DM annihilation, then we need our beyond SM physics to be able to 
explain this excess along with the observed DM relic density as well 
as neutrino masses and mixing. 
In this work we showed that the gauged $\bl$ extension of 
the Standard Model, which can very naturally explain lepton number 
violation and hence the existence of small 
Majorana neutrino masses, can also be extended to explain the relic DM 
density and the Fermi-LAT gamma ray excess, without conflicting 
with any existing theoretical or observational constraint. 

Since $\bl$ symmetry that we impose is local, there 
is an additional gauge boson $Z_{BL}$ in this model.
Three right handed neutrinos also have to included in the model to make it  
anomaly free. In order to break the $\bl$ symmetry spontaneously, one 
introduces an extra SM singlet scalar $\phi_H$ which carries a nontrivial ${\rm B-L}$
charge. The ${\rm B-L}$ charge of this scalar can be arranged in such a way 
that the right handed neutrinos pick up Majorana masses when $\phi_H$ 
gets a VEV, breaking the $\bl$ symmetry spontaneously. As a result the $Z_{BL}$ 
gauge boson also becomes massive. This extra neutral gauge boson has been 
searched for at collider experiments which put a stringent bound on the combination 
of the new $\bl$ gauge coupling and the mass of $Z_{BL}$. We extended this 
gauged $\bl$ model further by adding another complex SM scalar $\phi_{DM}$ 
which is charged under $\bl$ and arranged its $\bl$ charge in such away that 
all decays of $\phi_{DM}$ are forbidden making it a stable DM candidate. 

We next studied the viability of this model in simultaneously explaining the 
relic DM density of the Universe as well as the GC gamma ray excess 
through the annihilation of $\phi_{DM}$ and $\phi_{DM}^\dagger$ into 
the $b \bar{b}$ channel, mediated by $h_1$, $h_2$ and $Z_{BL}$, where 
$h_1$ and $h_2$ are the two neutral scalars in our model, of which we 
identified $h_1$ as the SM-like Higgs with its mass fixed at 125.5 GeV. 
There are 12 unknown new parameters in this model. 
We imposed constraints coming from vacuum stability, LEP bound on 
$M_{Z_{BL}}/g_{BL}$, LHC bounds on signal strength of the SM-like 
Higgs and invisible decay width of the SM-like Higgs, and found the regions 
of the model parameter space which can simultaneously explain the 
observed DM relic density as well as the Fermi-LAT GC gamma ray excess and 
at the same time evaded the bounds from the direct detection experiments 
such as LUX. We showed that for DM masses in the range 40-55 GeV
and for a wide range of $U(1)_{\rm B-L}$ gauge boson masses, one can
satisfy all these constraints if the additional neutral Higgs scalar
has a mass around the resonance region. We presented allowed areas in
the model parameter space consistent with all relevant data, calculated
the predicted gamma ray flux from the GC and discussed the related phenomenology. 

In conclusion, the observation of neutrino masses and dark matter are the two 
main observational evidences of physics beyond the Standard Model. The small 
neutrino masses can be explained in terms of lepton number violation and 
the gauged $U(1)_{\rm B-L}$ extension of the Standard Model can very easily 
accommodate the type-I seesaw mechanism since it must have three additional 
right-handed neutrinos for anomaly cancellation and
light Majorana neutrino masses 
are generated through this seesaw mechanism 
when the $U(1)_{\rm B-L}$ is broken spontaneously. We propose an 
extension of this model by adding a complex scalar field $\phi_{DM}$ that acts 
as the dark matter candidate and make it stable by arranging suitably its $\bl$ charge. 
We showed that both the relic abundance and the Galactic Centre gamma ray excess 
could be explained in this model. 
Thus this model can simultaneously explain neutrino masses,
the dark matter relic density of the Universe and the GC
gamma ray excess, while simultaneously satisfying all 
other collider and direct detection constraints.
This model should be testable at the next-generation
XENON1T experiment. 

\section*{Acknowledgment}
SK wishes to thank N. Chakrabarty and A. Gupta for discussions and
S. Kadge and M. Masud for help with numerical codes. The authors would
like to thank the Neutrino Project under the XII plan of Harish-Chandra
Research Institute. 
SK and AB also acknowledge the cluster computing facility at HRI
(http://cluster.hri.res.in).
This project has received funding from the European Union's
Horizon 2020 research and innovation programme under the
Marie Sklodowska-Curie grant agreement No 674896.

\label{acknowledge}


\begin{thebibliography}{99}
\bibitem{Sofue:2000jx} 
  Y.~Sofue and V.~Rubin,
  ``{\it Rotation curves of spiral galaxies}'',
  Ann.\ Rev.\ Astron.\ Astrophys.\  {\bf 39}, 137 (2001)
  [astro-ph/0010594].
\bibitem{Bartelmann:1999yn} 
  M.~Bartelmann and P.~Schneider,
  ``{\it Weak gravitational lensing}'',
  Phys.\ Rept.\  {\bf 340}, 291 (2001)
  [astro-ph/9912508].
\bibitem{Clowe:2003tk} 
  D.~Clowe, A.~Gonzalez and M.~Markevitch,
  ``{\it Weak lensing mass reconstruction of the interacting
  cluster 1E0657-558: Direct evidence for the existence of dark matter}'',
  Astrophys.\ J.\  {\bf 604}, 596 (2004)
  [astro-ph/0312273].
\bibitem{Hinshaw:2012aka} 
  G.~Hinshaw {\it et al.} [WMAP Collaboration],
  ``{\it Nine-Year Wilkinson Microwave Anisotropy Probe (WMAP)
  Observations: Cosmological Parameter Results}'',
  Astrophys.\ J.\ Suppl.\  {\bf 208}, 19 (2013)
  [arXiv:1212.5226 [astro-ph.CO]].
  \bibitem{Ade:2015xua} 
  P.~A.~R.~Ade {\it et al.} [Planck Collaboration],
  ``{\it Planck 2015 results. XIII. Cosmological parameters}'',
  arXiv:1502.01589 [astro-ph.CO].
\bibitem{Jungman:1995df} 
  G.~Jungman, M.~Kamionkowski and K.~Griest,
  ``{\it Supersymmetric dark matter}'',
  Phys.\ Rept.\  {\bf 267}, 195 (1996)
  [hep-ph/9506380].
\bibitem{Bertone:2004pz} 
  G.~Bertone, D.~Hooper and J.~Silk,
  ``{\it Particle dark matter: Evidence, candidates and constraints''},
  Phys.\ Rept.\  {\bf 405}, 279 (2005)
  [hep-ph/0404175].
\bibitem{McDonald:1993ex} 
  J.~McDonald,
  ``{\it Gauge singlet scalars as cold dark matter}'',
  Phys.\ Rev.\ D {\bf 50}, 3637 (1994)
  [hep-ph/0702143 [HEP-PH]].
\bibitem{Burgess:2000yq} 
  C.~P.~Burgess, M.~Pospelov and T.~ter Veldhuis,
  ``{\it The Minimal model of nonbaryonic dark matter: A Singlet scalar}''
  Nucl.\ Phys.\ B {\bf 619}, 709 (2001)
  [hep-ph/0011335].
\bibitem{Biswas:2011td} 
  A.~Biswas and D.~Majumdar,
  ``{\it The Real Gauge Singlet Scalar Extension of Standard Model:
  A Possible Candidate of Cold Dark Matter}'',
  Pramana {\bf 80}, 539 (2013)
  [arXiv:1102.3024 [hep-ph]].
\bibitem{Barbieri:2006dq} 
  R.~Barbieri, L.~J.~Hall and V.~S.~Rychkov,
  ``{\it Improved naturalness with a heavy Higgs: An Alternative
  road to LHC physics}'',
  Phys.\ Rev.\ D {\bf 74}, 015007 (2006)
  [hep-ph/0603188].
\bibitem{LopezHonorez:2006gr} 
  L.~Lopez Honorez, E.~Nezri, J.~F.~Oliver and M.~H.~G.~Tytgat,
  ``{\it The Inert Doublet Model: An Archetype for Dark Matter}'',
  JCAP {\bf 0702}, 028 (2007)
  [hep-ph/0612275].
\bibitem{Kim:2008pp} 
  Y.~G.~Kim, K.~Y.~Lee and S.~Shin,
  ``{\it Singlet fermionic dark matter}''
  JHEP {\bf 0805}, 100 (2008)
  [arXiv:0803.2932 [hep-ph]].

\bibitem{Kim:2016csm} 
  Y.~G.~Kim, K.~Y.~Lee, C.~B.~Park and S.~Shin,
  ``{\it Secluded singlet fermionic dark matter driven by the Fermi gamma-ray excess}'',
  Phys.\ Rev.\ D {\bf 93}, no. 7, 075023 (2016)
  [arXiv:1601.05089 [hep-ph]].

\bibitem{Hambye:2008bq} 
  T.~Hambye,
  ``{\it Hidden vector dark matter}'',
  JHEP {\bf 0901}, 028 (2009)
  [arXiv:0811.0172 [hep-ph]].
\bibitem{Lebedev:2011iq} 
  O.~Lebedev, H.~M.~Lee and Y.~Mambrini,
  ``{\it Vector Higgs-portal dark matter and the invisible Higgs}'',
  Phys.\ Lett.\ B {\bf 707}, 570 (2012)
  [arXiv:1111.4482 [hep-ph]].
\bibitem{Cao:2007fy} 
  Q.~H.~Cao, E.~Ma, J.~Wudka and C.-P.~Yuan,
  ``{\it Multipartite dark matter}'',
  arXiv:0711.3881 [hep-ph].
\bibitem{Zurek:2008qg} 
  K.~M.~Zurek,
  ``{\it Multi-Component Dark Matter}'',
  Phys.\ Rev.\ D {\bf 79}, 115002 (2009)
  [arXiv:0811.4429 [hep-ph]].
\bibitem{Belanger:2011ww} 
  G.~Belanger and J.~C.~Park,
  ``{\it Assisted freeze-out}''
  JCAP {\bf 1203}, 038 (2012)
  [arXiv:1112.4491 [hep-ph]].
\bibitem{Biswas:2013nn} 
  A.~Biswas, D.~Majumdar, A.~Sil and P.~Bhattacharjee,
  ``{\it Two Component Dark Matter : A Possible Explanation of
  130 GeV $\gamma-$ Ray Line from the Galactic Centre}'',
  JCAP {\bf 1312}, 049 (2013)
  [arXiv:1301.3668 [hep-ph]].
\bibitem{Bian:2013wna} 
  L.~Bian, R.~Ding and B.~Zhu,
  ``{\it Two Component Higgs-Portal Dark Matter}'',
  Phys.\ Lett.\ B {\bf 728}, 105 (2014)
  [arXiv:1308.3851 [hep-ph]].
\bibitem{Bhattacharya:2013hva} 
  S.~Bhattacharya, A.~Drozd, B.~Grzadkowski and J.~Wudka,
  ``{\it Two-Component Dark Matter}'',
  JHEP {\bf 1310}, 158 (2013)
  [arXiv:1309.2986 [hep-ph]].
\bibitem{Gondolo:1990dk} 
  P.~Gondolo and G.~Gelmini,
  ``{\it Cosmic abundances of stable particles: Improved analysis}'',
  Nucl.\ Phys.\ B {\bf 360}, 145 (1991).
\bibitem{Srednicki:1988ce} 
  M.~Srednicki, R.~Watkins and K.~A.~Olive,
  ``{\it Calculations of Relic Densities in the Early Universe}'',
  Nucl.\ Phys.\ B {\bf 310}, 693 (1988).
\bibitem{Akerib:2013tjd} 
  D.~S.~Akerib {\it et al.} [LUX Collaboration],
  ``{\it First results from the LUX dark matter experiment at the Sanford
  Underground Research Facility}'',
  Phys.\ Rev.\ Lett.\  {\bf 112}, 091303 (2014)
  [arXiv:1310.8214 [astro-ph.CO]].
\bibitem{Akerib:2015rjg} 
  D.~S.~Akerib {\it et al.} [LUX Collaboration],
  ``{\it Improved Limits on Scattering of Weakly Interacting
  Massive Particles from Reanalysis of 2013 LUX Data}'',
  Phys.\ Rev.\ Lett.\  {\bf 116}, no. 16, 161301 (2016)
  [arXiv:1512.03506 [astro-ph.CO]].
\bibitem{Aprile:2015uzo} 
  E.~Aprile {\it et al.} [XENON Collaboration],
  ``{\it Physics reach of the XENON1T dark matter experiment}'',
  JCAP {\bf 1604}, no. 04, 027 (2016)
  [arXiv:1512.07501 [physics.ins-det]].

\bibitem{Agnese:2014aze} 
  R.~Agnese {\it et al.} [SuperCDMS Collaboration],
  ``{\it Search for Low-Mass Weakly Interacting Massive
  Particles with SuperCDMS}'',
  Phys.\ Rev.\ Lett.\  {\bf 112}, no. 24, 241302 (2014)
  [arXiv:1402.7137 [hep-ex]].
\bibitem{Hooper:2009zm} 
  D.~Hooper,
  ``{\it Particle Dark Matter}'',
  arXiv:0901.4090 [hep-ph].
\bibitem{Goodenough:2009gk}
L.~Goodenough and D.~Hooper, 
``{\it Possible Evidence For Dark Matter
Annihilation In The Inner Milky Way From The Fermi Gamma Ray Space Telescope}'',
[arXiv:0910.2998 [hep-ph]].
\bibitem{Hooper:2010mq}
D.~Hooper and L.~Goodenough,
``{\it Dark Matter Annihilation in The Galactic
Center As Seen by the Fermi Gamma Ray Space Telescope}'',
{Phys.Lett}. {\bf B697} (2011) 412--428
[arXiv:1010.2752 [hep-ph]].
\bibitem{Boyarsky:2010dr}
A.~Boyarsky, D.~Malyshev, and O.~Ruchayskiy,
``{\it A comment on the emission
from the Galactic Center as seen by the Fermi telescope}'',
{Phys.Lett.} {\bf B705} (2011) 165--169
[arXiv:1012.5839 [hep-ph]].
\bibitem{Hooper:2011ti}
D.~Hooper and T.~Linden,
``{\it On The Origin Of The Gamma Rays From The Galactic Center}'',
{Phys.Rev.} {\bf D84} (2011) 123005
[arXiv:1110.0006 [astro-ph.HE]].
\bibitem{Abazajian:2012pn}
K.~N. Abazajian and M.~Kaplinghat,
``{\it Detection of a Gamma-Ray Source in the
Galactic Center Consistent with Extended Emission from Dark Matter
Annihilation and Concentrated Astrophysical Emission}'',
{Phys.Rev.} {\bf D86} (2012) 083511
[arXiv:1207.6047 [astro-ph.HE]].
\bibitem{Hooper:2013rwa}
D.~Hooper and T.~R. Slatyer,
``{\it Two Emission Mechanisms in the Fermi Bubbles:
A Possible Signal of Annihilating Dark Matter}'',
{Phys.Dark Univ.} {\bf 2} (2013) 118--138
[arXiv:1302.6589 [astro-ph.HE]].
\bibitem{Abazajian:2014fta}
K.~N. Abazajian, N.~Canac, S.~Horiuchi, and M.~Kaplinghat,
``{\it Astrophysical and Dark Matter Interpretations of Extended
Gamma-Ray Emission from the Galactic Center}'',
{Phys.Rev.} {\bf D90} (2014) 023526
[arXiv:1402.4090 [astro-ph.HE]].
\bibitem{Daylan:2014rsa} 
  T.~Daylan, D.~P.~Finkbeiner, D.~Hooper, T.~Linden, S.~K.~N.~Portillo,
  N.~L.~Rodd and T.~R.~Slatyer,
  ``{\it The characterization of the gamma-ray signal from the
  central Milky Way: A case for annihilating dark matter}'',
  Phys.\ Dark Univ.\  {\bf 12}, 1 (2016)
  [arXiv:1402.6703 [astro-ph.HE]].
\bibitem{TheFermi-LAT:2015kwa} 
  M.~Ajello {\it et al.} [Fermi-LAT Collaboration],
  ``{\it Fermi-LAT Observations of High-Energy $\gamma$-Ray Emission Toward the Galactic Center}'',
  Astrophys.\ J.\  {\bf 819}, no. 1, 44 (2016)
  [arXiv:1511.02938 [astro-ph.HE]].
\bibitem{Agrawal:2014oha} 
  P.~Agrawal, B.~Batell, P.~J.~Fox and R.~Harnik,
  ``{\it WIMPs at the Galactic Center}'',
  JCAP {\bf 1505}, 011 (2015)
  [arXiv:1411.2592 [hep-ph]].
\bibitem{Calore:2014nla} 
  F.~Calore, I.~Cholis, C.~McCabe and C.~Weniger,
  ``{\it A Tale of Tails: Dark Matter Interpretations of the Fermi GeV Excess
  in Light of Background Model Systematics}'',
  Phys.\ Rev.\ D {\bf 91}, no. 6, 063003 (2015)
  [arXiv:1411.4647 [hep-ph]].
\cite{Atwood:2009ez}
\bibitem{Atwood:2009ez} 
  W.~B.~Atwood {\it et al.} [Fermi-LAT Collaboration],
  ``{\it The Large Area Telescope on the Fermi Gamma-ray Space Telescope Mission}'',
  Astrophys.\ J.\  {\bf 697}, 1071 (2009)
  [arXiv:0902.1089 [astro-ph.IM]].
\bibitem{Lee:2015fea} 
  S.~K.~Lee, M.~Lisanti, B.~R.~Safdi, T.~R.~Slatyer and W.~Xue,
  ``{\it Evidence for Unresolved $\gamma$-Ray Point Sources in the Inner Galaxy}'',
  Phys.\ Rev.\ Lett.\  {\bf 116}, no. 5, 051103 (2016)
  [arXiv:1506.05124 [astro-ph.HE]].
\bibitem{Bartels:2015aea} 
  R.~Bartels, S.~Krishnamurthy and C.~Weniger,
  ``{\it Strong support for the millisecond pulsar origin of the Galactic
  center GeV excess}'',
  Phys.\ Rev.\ Lett.\  {\bf 116}, no. 5, 051102 (2016)
  [arXiv:1506.05104 [astro-ph.HE]].
\bibitem{Navarro:1996gj} 
  J.~F.~Navarro, C.~S.~Frenk and S.~D.~M.~White,
  ``{\it A Universal density profile from hierarchical clustering}'',
  Astrophys.\ J.\  {\bf 490}, 493 (1997)
  [astro-ph/9611107].
\bibitem{Boucenna:2011hy} 
  M.~S.~Boucenna and S.~Profumo,
  ``{\it Direct and Indirect Singlet Scalar Dark Matter Detection in the Lepton-Specific
  two-Higgs-doublet Model}'',
  Phys.\ Rev.\ D {\bf 84}, 055011 (2011)
  [arXiv:1106.3368 [hep-ph]].
\bibitem{Alvares:2012qv} 
  J.~D.~Ruiz-Alvarez, C.~A.~de S.Pires, F.~S.~Queiroz, D.~Restrepo and P.~S.~Rodrigues da Silva,
  ``{\it On the Connection of Gamma-Rays, Dark Matter and Higgs Searches at LHC}'',
  Phys.\ Rev.\ D {\bf 86}, 075011 (2012)
  [arXiv:1206.5779 [hep-ph]].
\bibitem{Okada:2013bna} 
  N.~Okada and O.~Seto,
  Phys.\ Rev.\ D {\bf 89}, no. 4, 043525 (2014)
  doi:10.1103/PhysRevD.89.043525
  [arXiv:1310.5991 [hep-ph]].
\bibitem{Alves:2014yha} 
  A.~Alves, S.~Profumo, F.~S.~Queiroz and W.~Shepherd,
  ``{\it Effective field theory approach to the Galactic Center gamma-ray excess}'',
  Phys.\ Rev.\ D {\bf 90}, no. 11, 115003 (2014)
  [arXiv:1403.5027 [hep-ph]].
  
\bibitem{Berlin:2014tja} 
  A.~Berlin, D.~Hooper and S.~D.~McDermott,
  ``{\it Simplified Dark Matter Models for the Galactic Center Gamma-Ray Excess}'',
  Phys.\ Rev.\ D {\bf 89}, 115022 (2014)
  [arXiv:1404.0022 [hep-ph]].
  \bibitem{Agrawal:2014una} 
  P.~Agrawal, B.~Batell, D.~Hooper and T.~Lin,
  ``{\it Flavored Dark Matter and the Galactic Center Gamma-Ray Excess}'',
  Phys.\ Rev.\ D {\bf 90}, 063512 (2014)
  [arXiv:1404.1373 [hep-ph]].
\bibitem{Izaguirre:2014vva} 
  E.~Izaguirre, G.~Krnjaic and B.~Shuve,
  ``{\it The Galactic Center Excess from the Bottom Up}'',
  Phys.\ Rev.\ D {\bf 90}, 055002 (2014)
  [arXiv:1404.2018 [hep-ph]].
\bibitem{Cerdeno:2014cda} 
  D.~G.~Cerdeño, M.~Peiró and S.~Robles,
  ``{\it Low-mass right-handed sneutrino dark matter: SuperCDMS and LUX
  constraints and the Galactic Centre gamma-ray excess}'',
  JCAP {\bf 1408}, 005 (2014)
  [arXiv:1404.2572 [hep-ph]].
\bibitem{Ipek:2014gua} 
  S.~Ipek, D.~McKeen and A.~E.~Nelson,
  ``{\it A Renormalizable Model for the Galactic Center Gamma Ray Excess from
  Dark Matter Annihilation}'',
  Phys.\ Rev.\ D {\bf 90}, 055021 (2014)
  [arXiv:1404.3716 [hep-ph]].
\bibitem{Boehm:2014bia} 
  C.~Boehm, M.~J.~Dolan and C.~McCabe,
  ``{\it A weighty interpretation of the Galactic Centre excess}'',
  Phys.\ Rev.\ D {\bf 90}, 023531 (2014)
  [arXiv:1404.4977 [hep-ph]].
\bibitem{Ko:2014gha} 
  P.~Ko, W.~I.~Park and Y.~Tang,
  ``{\it Higgs portal vector dark matter for $\mathinner{\mathrm{GeV}}$ scale
  $\gamma$-ray excess from galactic center}'',
  JCAP {\bf 1409}, 013 (2014)
  [arXiv:1404.5257 [hep-ph]].
\bibitem{Abdullah:2014lla} 
  M.~Abdullah, A.~DiFranzo, A.~Rajaraman, T.~M.~P.~Tait, P.~Tanedo and A.~M.~Wijangco,
  ``{\it Hidden on-shell mediators for the Galactic Center $\gamma$-ray excess}'',
  Phys.\ Rev.\ D {\bf 90}, no. 3, 035004 (2014)
  [arXiv:1404.6528 [hep-ph]].
\bibitem{Ghosh:2014pwa} 
  D.~K.~Ghosh, S.~Mondal and I.~Saha,
  ``{\it Confronting the Galactic Center Gamma Ray Excess With a Light Scalar Dark Matter}'',
  JCAP {\bf 1502}, no. 02, 035 (2015)
  [arXiv:1405.0206 [hep-ph]].
\bibitem{Martin:2014sxa} 
  A.~Martin, J.~Shelton and J.~Unwin,
  ``{\it Fitting the Galactic Center Gamma-Ray Excess with Cascade Annihilations}'',
  Phys.\ Rev.\ D {\bf 90}, no. 10, 103513 (2014)
  [arXiv:1405.0272 [hep-ph]].
\bibitem{Wang:2014elb} 
  L.~Wang and X.~F.~Han,
  ``{\it A simplified 2HDM with a scalar dark matter and the galactic center gamma-ray excess}'',
  Phys.\ Lett.\ B {\bf 739}, 416 (2014)
  [arXiv:1406.3598 [hep-ph]].
\bibitem{Basak:2014sza} 
  T.~Mondal and T.~Basak,
  ``{\it Class of Higgs-portal Dark Matter models in the light of gamma-ray excess
  from Galactic center}'',
  Phys.\ Lett.\ B {\bf 744}, 208 (2015)
  [arXiv:1405.4877 [hep-ph]].
\bibitem{Detmold:2014qqa} 
  W.~Detmold, M.~McCullough and A.~Pochinsky,
  ``{\it Dark Nuclei I: Cosmology and Indirect Detection}'',
  Phys.\ Rev.\ D {\bf 90}, 115013 (2014)
  [arXiv:1406.2276 [hep-ph]].
\bibitem{Arina:2014yna} 
  C.~Arina, E.~Del Nobile and P.~Panci,
  ``{\it Dark Matter with Pseudoscalar-Mediated Interactions Explains
  the DAMA Signal and the Galactic Center Excess}'',
  Phys.\ Rev.\ Lett.\  {\bf 114}, 011301 (2015)
  [arXiv:1406.5542 [hep-ph]].
\bibitem{Okada:2014usa} 
  N.~Okada and O.~Seto,
  ``{\it Galactic Center gamma-ray excess from two-Higgs-doublet-portal dark matter}'',
  Phys.\ Rev.\ D {\bf 90}, no. 8, 083523 (2014)
  [arXiv:1408.2583 [hep-ph]].
\bibitem{Ghorbani:2014qpa} 
  K.~Ghorbani,
  ``{\it Fermionic dark matter with pseudo-scalar Yukawa interaction}'',
  JCAP {\bf 1501}, 015 (2015)
  [arXiv:1408.4929 [hep-ph]].
\bibitem{Banik:2014eda} 
  A.~D.~Banik and D.~Majumdar,
  ``{\it Low Energy Gamma Ray Excess Confronting a Singlet Scalar Extended
  Inert Doublet Dark Matter Model}'',
  Phys.\ Lett.\ B {\bf 743}, 420 (2015)
  [arXiv:1408.5795 [hep-ph]].
\bibitem{Biswas:2014hoa} 
  A.~Biswas,
  ``{\it Explaining Low Energy $\gamma$-ray Excess from the Galactic
  Centre using a Two Component Dark Matter Model}'',
  J.\ Phys.\ G {\bf 43}, no. 5, 055201 (2016)
  [arXiv:1412.1663 [hep-ph]].
\bibitem{Ghorbani:2014gka} 
  K.~Ghorbani and H.~Ghorbani,
  ``{\it Scalar split WIMPs in future direct detection experiments}'',
  Phys.\ Rev.\ D {\bf 93}, no. 5, 055012 (2016)
  [arXiv:1501.00206 [hep-ph]].
\bibitem{Cerdeno:2015ega} 
  D.~G.~Cerdeno, M.~Peiro and S.~Robles,
  ``{\it Fits to the Fermi-LAT GeV excess with RH sneutrino dark matter:
  implications for direct and indirect dark matter searches and the LHC}'',
  Phys.\ Rev.\ D {\bf 91}, no. 12, 123530 (2015)
  [arXiv:1501.01296 [hep-ph]].
\bibitem{Biswas:2015sva} 
  A.~Biswas, D.~Majumdar and P.~Roy,
  ``{\it Nonthermal two component dark matter model for Fermi-LAT
  $\gamma$-ray excess and 3.55 keV X-ray line}'',
  JHEP {\bf 1504}, 065 (2015)
  [arXiv:1501.02666 [hep-ph]].
\bibitem{Caron:2015wda} 
  A.~Achterberg, S.~Amoroso, S.~Caron, L.~Hendriks, R.~Ruiz de Austri and C.~Weniger,
  ``{\it A description of the Galactic Center excess in the Minimal
  Supersymmetric Standard Model}'',
  JCAP {\bf 1508}, no. 08, 006 (2015)
  [arXiv:1502.05703 [hep-ph]].
\bibitem{Borah:2015rla} 
  D.~Borah, A.~Dasgupta and R.~Adhikari,
  ``{\it Common origin of the 3.55 keV x-ray line and the Galactic
  Center gamma-ray excess in a radiative neutrino mass model}'',
  Phys.\ Rev.\ D {\bf 92}, no. 7, 075005 (2015)
  [arXiv:1503.06130 [hep-ph]].
\bibitem{Banik:2015aya} 
  A.~D.~Banik, D.~Majumdar and A.~Biswas,
  ``{\it Possible Explanation of Indirect Gamma Ray
  Signatures from Hidden Sector SU(2)$_{\rm H}$ Fermionic Dark Matter}'',
  arXiv:1506.05665 [hep-ph].
\bibitem{Dutta:2015ysa} 
  B.~Dutta, Y.~Gao, T.~Ghosh and L.~E.~Strigari,
  ``{\it Confronting Galactic center and dwarf spheroidal
  gamma-ray observations with cascade annihilation models}'',
  Phys.\ Rev.\ D {\bf 92}, no. 7, 075019 (2015)
  [arXiv:1508.05989 [hep-ph]].
  
\bibitem{Cuoco:2016jqt} 
  A.~Cuoco, B.~Eiteneuer, J.~Heisig and M.~Krämer,
  ``{\it A global fit of the $\gamma$-ray galactic center excess within the scalar singlet Higgs portal model}'',
  arXiv:1603.08228 [hep-ph].
  
\bibitem{Mohapatra:1980qe} 
  R.~N.~Mohapatra and R.~E.~Marshak,
  ``{\it Local B-L Symmetry of Electroweak Interactions, Majorana Neutrinos and Neutron Oscillations}'',
  Phys.\ Rev.\ Lett.\  {\bf 44}, 1316 (1980)
  Erratum: [Phys.\ Rev.\ Lett.\  {\bf 44}, 1643 (1980)].
\bibitem{Georgi:1981pg} 
  H.~M.~Georgi, S.~L.~Glashow and S.~Nussinov,
  ``{\it Unconventional Model of Neutrino Masses}'',
  Nucl.\ Phys.\ B {\bf 193}, 297 (1981).
  
\bibitem{Wetterich:1981bx} 
  C.~Wetterich,
  ``{\it Neutrino Masses and the Scale of B-L Violation}'',
  Nucl.\ Phys.\ B {\bf 187}, 343 (1981).

\bibitem{Lindner:2011it} 
  M.~Lindner, D.~Schmidt and T.~Schwetz,
  ``{\it Dark Matter and neutrino masses from global U(1)$_{B-L}$ symmetry breaking}'',
  Phys.\ Lett.\ B {\bf 705}, 324 (2011)
  [arXiv:1105.4626 [hep-ph]].

\bibitem{Okada:2010wd} 
  N.~Okada and O.~Seto,
  ``{\it Higgs portal dark matter in the minimal gauged $U(1)_{B-L}$ model}'',
  Phys.\ Rev.\ D {\bf 82}, 023507 (2010)
  [arXiv:1002.2525 [hep-ph]].

\bibitem{Okada:2012sg} 
  N.~Okada and Y.~Orikasa,
  ``{\it Dark matter in the classically conformal B-L model}'',
  Phys.\ Rev.\ D {\bf 85}, 115006 (2012)
  [arXiv:1202.1405 [hep-ph]].
  
\bibitem{Basso:2012ti} 
  L.~Basso, O.~Fischer and J.~J.~van der Bij,
  ``{\it Natural Z$^{\prime}$ model with an inverse seesaw mechanism and leptonic dark matter}'',
  Phys.\ Rev.\ D {\bf 87}, no. 3, 035015 (2013)
  [arXiv:1207.3250 [hep-ph]].

\bibitem{Basak:2013cga} 
  T.~Basak and T.~Mondal,
  ``{\it Constraining Minimal $U(1)_{B-L}$ model from Dark Matter Observations}'',
  Phys.\ Rev.\ D {\bf 89}, 063527 (2014)
  [arXiv:1308.0023 [hep-ph]].
  
\bibitem{Sanchez-Vega:2014rka} 
  B.~L.~Sánchez-Vega, J.~C.~Montero and E.~R.~Schmitz,
  ``{\it Complex Scalar DM in a B-L Model}'',
  Phys.\ Rev.\ D {\bf 90}, no. 5, 055022 (2014)
  [arXiv:1404.5973 [hep-ph]].
  
  
\bibitem{Guo:2015lxa} 
  J.~Guo, Z.~Kang, P.~Ko and Y.~Orikasa,
  ``{\it Accidental dark matter: Case in the scale invariant local B-L model}'',
  Phys.\ Rev.\ D {\bf 91}, no. 11, 115017 (2015)
  [arXiv:1502.00508 [hep-ph]].


\bibitem{Rodejohann:2015lca} 
  W.~Rodejohann and C.~E.~Yaguna,
  ``{\it Scalar dark matter in the ${\rm B-L}$ model}'',
  JCAP {\bf 1512}, no. 12, 032 (2015)
  [arXiv:1509.04036 [hep-ph]].
  
\bibitem{Okada:2016gsh} 
  N.~Okada and S.~Okada,
  ``{\it $Z^\prime_{BL}$ portal dark matter and LHC Run-2 results}'',
  Phys.\ Rev.\ D {\bf 93}, no. 7, 075003 (2016)
  [arXiv:1601.07526 [hep-ph]].
 
\bibitem{Buchmuller:1992qc} 
  W.~Buchmuller and T.~Yanagida,
  ``{\it Baryogenesis and the scale of B-L breaking}'',
  Phys.\ Lett.\ B {\bf 302}, 240 (1993).
 
\bibitem{Buchmuller:1996pa} 
  W.~Buchmuller and M.~Plumacher,
  ``{\it Baryon asymmetry and neutrino mixing}'',
  Phys.\ Lett.\ B {\bf 389}, 73 (1996)
  [hep-ph/9608308].
  
\bibitem{Dulaney:2010dj} 
  T.~R.~Dulaney, P.~Fileviez Perez and M.~B.~Wise,
  ``{\it Dark Matter, Baryon Asymmetry, and Spontaneous B and L Breaking}'',
  Phys.\ Rev.\ D {\bf 83}, 023520 (2011)
  [arXiv:1005.0617 [hep-ph]].
\bibitem{Carena:2004xs} 
  M.~Carena, A.~Daleo, B.~A.~Dobrescu and T.~M.~P.~Tait,
  ``{\it $Z^\prime$ gauge bosons at the Tevatron}'',
  Phys.\ Rev.\ D {\bf 70}, 093009 (2004)
  [hep-ph/0408098].
\bibitem{Cacciapaglia:2006pk} 
  G.~Cacciapaglia, C.~Csaki, G.~Marandella and A.~Strumia,
  ``{\it The Minimal Set of Electroweak Precision Parameters}'',
  Phys.\ Rev.\ D {\bf 74}, 033011 (2006)
  [hep-ph/0604111].
\bibitem{Aad:2014cka} 
  G.~Aad {\it et al.} [ATLAS Collaboration],
  Phys.\ Rev.\ D {\bf 90}, no. 5, 052005 (2014)
  doi:10.1103/PhysRevD.90.052005
  [arXiv:1405.4123 [hep-ex]].
\bibitem{Cline:2013gha} 
  J.~M.~Cline, K.~Kainulainen, P.~Scott and C.~Weniger,
  ``{\it Update on scalar singlet dark matter}'',
  Phys.\ Rev.\ D {\bf 88}, 055025 (2013)
  Erratum: [Phys.\ Rev.\ D {\bf 92}, no. 3, 039906 (2015)]
  [arXiv:1306.4710 [hep-ph]].
\bibitem{Belanger:2008sj} 
  G.~Belanger, F.~Boudjema, A.~Pukhov and A.~Semenov,
  ``{\it Dark matter direct detection rate in a generic model with micrOMEGAs 2.2}'',
  Comput.\ Phys.\ Commun.\  {\bf 180}, 747 (2009)
  [arXiv:0803.2360 [hep-ph]].
\bibitem{Agashe:2014kda} 
  K.~A.~Olive {\it et al.} [Particle Data Group Collaboration],
  ``{\it Review of Particle Physics}'',
  Chin.\ Phys.\ C {\bf 38}, 090001 (2014).
\bibitem{Bechtle:2014ewa} 
  P.~Bechtle, S.~Heinemeyer, O.~Stål, T.~Stefaniak and G.~Weiglein,
  ``{\it Probing the Standard Model with Higgs signal rates from the
  Tevatron, the LHC and a future ILC}'',
  JHEP {\bf 1411}, 039 (2014)
  [arXiv:1403.1582 [hep-ph]].
\bibitem{Belanger:2013kya} 
  G.~Belanger, B.~Dumont, U.~Ellwanger, J.~F.~Gunion and S.~Kraml,
  ``{\it Status of invisible Higgs decays}'',
  Phys.\ Lett.\ B {\bf 723}, 340 (2013)
  [arXiv:1302.5694 [hep-ph]].
\bibitem{Edsjo:1997bg} 
  J.~Edsjo and P.~Gondolo,
  ``{\it Neutralino relic density including coannihilations}'',
  Phys.\ Rev.\ D {\bf 56}, 1879 (1997)
  [hep-ph/9704361].
\bibitem{Belanger:2013oya} 
  G.~Belanger, F.~Boudjema, A.~Pukhov and A.~Semenov,
  ``{\it micrOMEGAs-3: A program for calculating dark matter observables}'',
  Comput.\ Phys.\ Commun.\  {\bf 185}, 960 (2014)
  [arXiv:1305.0237 [hep-ph]].
\bibitem{Semenov:2010qt} 
  A.~Semenov,
  ``{\it LanHEP - a package for automatic generation of Feynman
  rules from the Lagrangian. Updated version 3.1}'',
  arXiv:1005.1909 [hep-ph].
\bibitem{Cirelli:2010xx} 
  M.~Cirelli {\it et al.},
  ``{\it PPPC 4 DM ID: A Poor Particle Physicist Cookbook for Dark Matter Indirect Detection}'',
  JCAP {\bf 1103}, 051 (2011)
  Erratum: [JCAP {\bf 1210}, E01 (2012)]
  [arXiv:1012.4515 [hep-ph]].
  \bibitem{einasto}
  J.~Einasto, 
  ``{\it On the construction of a composite model for the galaxy and on
  the determination of the system of galactic parameters}'',
  Trudy Inst. Astrofiz. Alma-Ata {\bf 5} 87 (1965).
\bibitem{Bergstrom:1997fj} 
  L.~Bergstrom, P.~Ullio and J.~H.~Buckley,
  ``{\it Observability of gamma-rays from dark matter
  neutralino annihilations in the Milky Way halo}'',
  Astropart.\ Phys.\  {\bf 9}, 137 (1998)
  [astro-ph/9712318].
\bibitem{Moore:1999gc} 
  B.~Moore, T.~R.~Quinn, F.~Governato, J.~Stadel and G.~Lake,
  ``{\it Cold collapse and the core catastrophe}'',
  Mon.\ Not.\ Roy.\ Astron.\ Soc.\  {\bf 310}, 1147 (1999)
  [astro-ph/9903164].
  \end{thebibliography}
\end{document}